# BARIUM ISOTOPIC COMPOSITION OF MAINSTREAM SILICON CARBIDES FROM MURCHISON: CONSTRAINTS FOR *s*-PROCESS NUCLEOSYNTHESIS IN AGB STARS

NAN LIU[1,2,3], MICHAEL R. SAVINA[2,3], ANDREW M. DAVIS[1,2,4], ROBERTO GALLINO[5], OSCAR STRANIERO[6], FRANK GYNGARD[7], MICHAEL J. PELLIN[1,2,3], DAVID G. WILLINGHAM[3], NICOLAS DAUPHAS[1,2,4], MARCO PIGNATARI[8,10], SARA BISTERZO[5,11], SERGIO CRISTALLO[6] AND FALK HERWIG[9,10].

[1]Department of the Geophysical Sciences, The University of Chicago, Chicago, IL 60637, USA; lnsmile@uchicago.edu;

[2]Chicago Center for Cosmochemistry, Chicago, IL 60637, USA;

[3]Materials Science Division, Argonne National Laboratory, Argonne, IL 60439, USA;

[4]Enrico Fermi Institute, The University of Chicago, Chicago, IL 60637, USA;

[5]Dipartimento di Fisica, Università di Torino, Torino I-10125, Italy;

[6]INAF−Osservatorio Astronomico di Collurania, via Maggini snc 64100, Teramo, Italy;

[7]Laboratory for Space Sciences, Washington University, St. Louis 63130, MO, USA;

[8]Department of Physics, University of Basel, Basel CH-4056, Switzerland;

[9]Department of Physics & Astronomy, University of Victoria, Victoria, BC V8P5C2, Canada;

[11]INAF−Osservatorio Astrofisico di Torino−Strada Osservatorio 20, Pino Torinese I-10025, Italy.

Short title: Barium isotopic composition of mainstream SiCs

[10] NuGrid collaboration, http://www.nugridstars.org.

## ABSTRACT

We present barium, carbon, and silicon isotopic compositions of 38 acid-cleaned presolar SiC grains from Murchison. Comparison with previous data shows that acid washing is highly effective in removing barium contamination. Strong depletions in $\delta(^{138}Ba/^{136}Ba)$ values are found, down to −400 ‰, which can only be modeled with a flatter $^{13}C$ profile within the $^{13}C$ pocket than is normally used. The dependence of $\delta(^{138}Ba/^{136}Ba)$ predictions on the distribution of $^{13}C$ within the pocket in AGB models allows us to probe the $^{13}C$ profile within the $^{13}C$ pocket and the pocket mass in asymptotic giant branch (AGB) stars. In addition, we provide constraints on the $^{22}Ne(\alpha,n)^{25}Mg$ rate in the stellar temperature regime relevant to AGB stars, based on $\delta(^{134}Ba/^{136}Ba)$ values of mainstream grains. We found two nominally mainstream grains with strongly negative $\delta(^{134}Ba/^{136}Ba)$ values that cannot be explained by any of the current AGB model calculations. Instead, such negative values are consistent with the intermediate neutron capture process (*i*-process), which is activated by the Very Late Thermal Pulse (VLTP) during the post-AGB phase and characterized by a neutron density much higher than the *s*-process. These two grains may have condensed around post-AGB stars. Finally, we report abundances of two *p*-process isotopes, $^{130}Ba$ and $^{132}Ba$, in single SiC grains. These isotopes are destroyed in the *s*-process in AGB stars. By comparing their abundances with respect to that of $^{135}Ba$, we conclude that there is no measurable decay of $^{135}Cs$ ($t_{½}$= 2.3 Ma) to $^{135}Ba$ in individual SiC grains, indicating condensation of barium, but not cesium into SiC grains before $^{135}Cs$ decayed.

## 1. INTRODUCTION

Presolar silicon carbides (SiC grains) are pristine microcrystals that condensed in carbon-rich stellar winds and/or explosions (Lodders & Fegley 1995), were ejected into the interstellar medium preserving their nucleosynthetic origin, transported to the protosolar nebula, incorporated in meteorite parent bodies, and delivered to Earth in meteorites, where they were discovered over 25 years ago via their exotic isotopic signatures (Bernatowicz et al. 1987; Zinner et al. 1987; Lewis et al. 1990). Extensive studies of isotopic anomalies of light elements (A<56) in presolar SiC grains by Secondary Ion Mass Spectrometry (SIMS) confirmed that different types of SiC grains have different types of parent stars, with the majority of them (mainstream grains) originating from low-mass Asymptotic Giant Branch (AGB) stars (Hoppe et al. 1994;

Zinner 2004; Clayton & Nittler 2004; Davis 2011). AGB stars are the astrophysical source of the main *s*-process component (Gallino et al. 1990; Arlandini et al. 1999; Bisterzo et al. 2011).

Previous isotopic measurements of heavy elements (strontium, ruthenium, zirconium, molybdenum and barium) in single mainstream grains by Resonance Ionization Mass Spectrometry (RIMS) showed clear *s*-process signatures (Nicolussi et al. 1997, 1998; Savina et al. 2003a, 2004; Barzyk et al. 2007a), providing constraints on free parameters in AGB nucleosynthesis calculations. Mainstream SiC grains condensed in the envelope outflows of AGB stars prior to solar system formation. The measurement of abundance anomalies in their isotopic compositions allows the study of *s*-process nucleosynthesis in individual stars at a level of precision unavailable to spectroscopic observations.

Barium isotopic compositions in presolar grains have been measured in SiC aggregates (Ott & Begemann 1990; Zinner et al. 1991; Prombo et al. 1993; Jennings et al. 2002) and in single mainstream grains (Jennings et al. 2002; Savina et al. 2003a; Barzyk et al. 2007a; Marhas et al. 2007; Ávila et al. 2013). With the exception of very large mainstream grains (7–58 $\mu m$, Ávila et al. 2013), *s*-process barium isotopic patterns were found, although there exist systematic differences between single grain data and model predictions (Lugaro et al. 2003a). In addition, it was found that mainstream SiC grains with smaller sizes tend to contain higher barium concentrations and a barium isotopic signature more strongly enriched in *s*-process isotopes (Zinner et al. 1991; Marhas et al. 2007; Ávila et al. 2013), supporting the view that barium was implanted in the grains. However, the implantation model proposed by Verchovsky et al. (2004) predicts that the 'G-component' (pure *s*-process barium isotopes made in the helium intershell), defined by Lewis et al. (1990), should be implanted more efficiently into larger SiC grains at high energy compared to the 'N-component' (initial barium isotopes present in the convective envelope, low energy), which is the opposite of the observed trend. More importantly, the implantation scenario would have cesium coimplanted with barium into SiC grains due to their similar ionic radii, masses, and ionization potentials. The fact that there is no detectable radiogenic $^{135}$Ba from $^{135}$Cs decay in SiC aggregates (Lugaro et al. 2003a) challenged the implantation scenario for the refractory element barium.

Barzyk et al. (2007a) found solar system barium contamination in Murchison mainstream SiC grains and proposed that the contamination was caused by aqueous alteration on the Murchison parent body. Clean samples are therefore a prerequisite in order to study the size-dependent trend of barium isotopic compositions in mainstream SiC grains.

While other heavy elements can suffer from severe isobaric interferences and can only be measured by RIMS, barium isotopes are the most abundant stable nuclides in the atomic mass region around A = 140 and are only interfered by the isotopes of xenon, which is present in presolar SiC in extremely low concentrations. Thus, barium isotopes can be measured using SIMS and Thermal Ionization Mass Spectrometry (TIMS). The previous barium isotopic data on SiC aggregates by TIMS and on single SiC grains by a high spatial resolution SIMS instrument, the NanoSIMS, however, suffered from potential molecular interferences, which complicated the interpretation of the results (Ávila et al. 2013). Early SIMS measurements on SiC aggregates and recent Sensitive High Resolution Ion Microprobe-Reverse Geometry (SHRIMP-RG) measurements on large mainstream SiC grains used energy filtering to suppress molecular interferences in the barium mass region (Zinner et al. 1991; Ávila et al. 2013). A few grains were analyzed using RIMS by Jennings et al. (2002) and Savina et al. (2003a). The precision of RIMS barium isotopic data obtained subsequently by Barzyk et al. (2007a) was limited by low counting statistics since they aimed to do multielement measurements sequentially in presolar grains, with barium measurements completed after much of the grain material had been consumed in measuring molybdenum and zirconium isotopic compositions. In the present work, we report barium isotopic measurements of 61 individual acid-washed presolar SiC grains from Murchison, out of which only 40 had high enough barium concentrations to give adequate counting statistics for a reliable estimate of isotopic ratios. The carbon and silicon isotopic compositions of most of these grains were subsequently measured by NanoSIMS.

One of the aims of this work was to minimize the Solar System barium contamination in Murchison mainstream SiC grains found by Barzyk et al. (2007a). We succeeded in making barium measurements in grains free from Solar System barium contamination, providing more powerful constraints for *s*-process nucleosynthesis in AGB stars.

## 2. EXPERIMENTAL METHODS

The SiC separation and mounting methods were described by Levine et al. (2009). Briefly, we followed the standard procedure (Amari et al. 1994) to extract presolar SiC grains from the Murchison meteorite. Grains were treated additionally with concentrated $H_2SO_4$ at 200°*C* for 12 *h*, $HClO_4$ at 195°*C* for 4 *h*, a solution of 10 *M* HF and 1.2 *M* HCl at 20°*C* for 12 *h*, and $HClO_4$ at 180°*C* for 3 *h* in order to remove any parent-body or terrestrial contamination. The SiC grains used in this study are from the KJG series and were separated in size by sedimentation (Amari et al. 1994); they are typically 1−3 *μm* in diameter. Some KJG grains were deposited on a high purity gold foil from a water-isopropanol suspension and pressed into the gold foil with a sapphire disk. SiC grains were identified on the mount with secondary electron and energy dispersive X-ray images prior to RIMS analysis. Most of the grains are well separated from each other (more than 20 *μm* apart).

Barium isotopic compositions of the presolar SiC grains were measured on the CHARISMA instrument at Argonne National Laboratory using the experimental methods described by Savina et al. (2003a, 2003b). The two-color resonance ionization scheme used for barium in this study was that of Barzyk et al. (2007a, 2007b), which is different from the one used by Savina et al. (2003a, 2003b). According to the saturation curves reported in Fig. 2 of Barzyk et al. (2007b), both resonance and ionization transitions are well saturated with the beam intensities used in this study. We used a rastered ~1 *μm*-size UV desorption laser (351 *nm*) to release material with a raster size sufficient to desorb from the complete grain. Rastered areas are 10×10 to 20×20 *μm* in size and are thus smaller than the distances between grains in almost all cases. Postanalysis imaging with a scanning electron microscope verified that no more than one grain was analyzed at a time. Twenty-one of the 61 grains analyzed had so little barium that the desorbing laser fluence required to produce a signal was damaging the gold mount and resulted in significant backgrounds due to secondary ions.

Updates to CHARISMA since the previous studies include new Ti:sapphire lasers that produce much more powerful beams with broader bandwidth in order to diminish the effect of isotope shifts, which had previously limited the precision of RIMS measurements (Isselhardt et al. 2011). Additionally, more powerful beams suppress odd/even isotope effects and therefore yield smaller isotopic fractionation between odd and even mass isotopes (Fairbank et al. 1997; Wunderlich et al. 1992, 1993).

After RIMS analysis, carbon and silicon isotopic compositions in the 40 grains were determined with the Cameca NanoSIMS 50 at Washington University, by rastering a primary $Cs^+$ beam over each grain and simultaneously collecting secondary ions of $^{12}C^-$, $^{13}C^-$, $^{28}Si^-$, $^{29}Si^-$, and $^{30}Si^-$. One of the 40 grains was determined to be of type AB and another to be of type Z (Hoppe et al. 1994, 1997). Of the remaining 38 grains, 24 grains were determined to be mainstream. The remaining 14 grains were completely consumed during the RIMS measurement and could not be classified by NanoSIMS; these are grouped as mainstream grains for purposes of discussion since >90 % of SiC grains are mainstream (Hoppe et al. 1994). Carbon, silicon and barium isotope ratios of the 38 mainstream and unclassified grains are reported in Table 1. All uncertainties are reported as 2σ, and include both counting errors and external reproducibility.

Carbon isotopic data are reported as ratios; silicon and barium isotopic data are given as δ-values, defined as deviation in parts per thousand from isotope ratios measured in samples relative to standards (e.g., $\delta^{134}Ba = [(^{134}Ba/^{136}Ba)_{grain}/(^{134}Ba/^{136}Ba)_{standard} - 1)\times 1000]$. For NanoSIMS measurements of carbon and silicon isotopes, terrestrial SiC aggregates were used as standards and measured in between every ten grain measurements in order to monitor and correct for instrumental drift. Carbon-12 and $^{28}Si$ are chosen as the reference isotopes for carbon and silicon, respectively. The uncertainties of NanoSIMS data are calculated by including errors from both counting statistics and the overall scatter on the measured standards. For the RIMS study, terrestrial $BaTiO_3$ was measured as a standard on each day prior to grain measurements. In previous RIMS studies (e.g., Barzyk et al. 2007a), data uncertainties were underestimated, as only uncertainties resulted from counting statistics were taken into consideration. More reliable data are obtained in this study by using Isoplot software (Ludwig 2012) to calculate Mean Square Weighted Deviations (MSWDs) of standard measurements in order to estimate uncertainties beyond Poisson statistics related to instabilities of the instrument and laser beams. No long-time drifting is found and MSWDs are close to unity, which demonstrates that instrumental instabilities are not a significant source of uncertainty. The error calculation equation for barium isotope ratios used in this study is given by

$$2\sigma = 2\times(\delta+1000)\times\sqrt{\mathrm{MSWD}\times(\frac{1}{^{i}\mathrm{Ba}_{grain}}+\frac{1}{^{136}\mathrm{Ba}_{grain}}+\frac{1}{^{i}\mathrm{Ba}_{std}}+\frac{1}{^{136}\mathrm{Ba}_{std}})} \qquad (1)$$

where $^{i}Ba_{grain}$, etc. are the number of atoms counted.

## 3. RESULTS

A partial section of the chart of the nuclides in the xenon-lanthanum region is shown in Fig. 1. Although $^{134}Ba$ and $^{136}Ba$ both are pure *s*-process isotopes shielded by their stable xenon isobars, their relative abundances produced during AGB nucleosynthesis may deviate from the solar ratio because of the branch point at $^{134}Cs$. The stellar $\beta^-$ decay rate of $^{134}Cs$ has a strong temperature dependence (Takahashi & Yokoi 1987, TY87 hereafter), increasing almost two orders of magnitude as the temperature rises to ~$3\times10^8$ *K* during thermal pulses (TPs). Despite its shorter half-life at higher stellar temperature, the relatively high peak neutron density during TPs due to the marginal activation of the $^{22}$Ne neutron source ($n_n$ = ~$10^9$ $cm^{-3}$, compared to $10^7$−$10^8$ $cm^{-3}$ for $^{13}$C neutron source) can increase the neutron capture rate of $^{134}$Cs above its $\beta^-$ decay rate, such that $^{135}$Cs production is favored over $^{134}$Ba production. Once $^{135}$Cs is produced, it is stable ($t_{1/2}$ = 2.3 *Ma*) on the timescale of the *s*-process in AGB stars (*t* ~ 20 *ka*, Gallino et al. 1998) and continues to undergo neutron capture to form unstable $^{136}$Cs ($t_{1/2}$ = 13 *days*), almost all of which decays to $^{136}$Ba. Thus high neutron fluxes partially bypass both $^{134}$Ba and $^{135}$Ba, and accumulate $^{136}$Ba. As shown in Fig. 1, these two branches in the *s*-process path in the cesium-barium region join at $^{136}$Ba, so little branching effect is seen for $^{137}$Ba or $^{138}$Ba (Lugaro et al. 2003a).

Barium isotopic compositions of the 38 grains, along with carbon and silicon data when available, are reported in Table 1, including 24 mainstream and 14 unclassified grains. Errors in Table 1 are given as 2σ. Seven mainstream grains had a significant number of counts of the rare *p*-only isotopes $^{130}$Ba and $^{132}$Ba (0.1 % abundance each in terrestrial barium, as shown in Fig. 1), so we summed the counts of $^{130}$Ba and $^{132}$Ba in order to reduce the uncertainties in δ-values. The data are reported as $\delta^{130+132}$Ba in Table 1 and are used to discuss AGB nucleosynthesis models and barium condensation into SiC grains around AGB stars.

All barium data are plotted in Fig. 2. In general, unclassified grains have relatively small error bars because we consumed the grains in their entirety and therefore had more barium counts. Nearly all of the grains have strongly negative $\delta^{135}$Ba values (< −400 ‰) with the exception of two unclassified grains (G379 and G89, which are shown as a blue triangle and a grey dot, respectively, in Fig. 2). The barium isotopic compositions of the mainstream and unclassified grains generally agree with previous studies and with AGB model predictions (see below) with a few exceptions. Two mainstream grains (G244 and G232, shown in red in Fig. 2 and highlighted in Table 1) have strongly negative $\delta^{134}$Ba values in comparison to both previous studies and AGB model predictions, and might require a different stellar source such as post-AGB stars (see discussion in Section 4.5).

We chose mainstream and unclassified grains from this and previous studies whose 2σ errors in $\delta^{135}$Ba were less than 160 ‰ (the number is chosen to include most of the grains from this and previous studies while excluding the ones with relatively large uncertainties) and plotted them in Fig. 3. This criterion was used to eliminate the effects of varying useful yields, and of different amounts of grains consumed and analyzed in these measurements. More importantly, grain data with less uncertainty allows us to derive more stringent constraints on stellar model predictions. Six of the 38 mainstream and unclassified grains from this study have 2σ errors greater than 160 ‰ and are therefore excluded in Fig. 3. Because the error is linear in the δ-value (Equation 1), larger errors can result from fewer counts and/or higher δ-values. This could introduce a selection bias against grains with higher δ-values. Only one of the six excluded grains (G89) plots significantly outside the cluster of known mainstream grains. This grain is unclassified as shown in Table 1; thus the criterion excludes grains primarily on the basis of lower barium counts and causes little or no selection bias with respect to isotope ratios.

### *3.1. Solar System Barium Contamination in SiC Grains*

The isotopic composition of presolar SiC grains contaminated with Solar System material is indistinguishable from the 'N-component' in AGB stellar envelopes when making comparisons between grain data and model predictions based on stars starting with near-solar metallicity (initial isotopic composition variations of model predictions are within ±200 ‰, Bisterzo et al. 2011). Models with low mean neutron exposures can account for near-solar barium isotopic compositions in the absence of contamination. Barzyk et al. (2007a) did multielement/multi-isotope analysis of single SiC grains and found that five of 23 Murchison grains were contaminated with Solar System barium. Marhas et al. (2007) imaged the spatial distribution of carbon, silicon and barium signals for each presolar SiC grain with NanoSIMS and found barium-rich rims around or on the surfaces of some of the grains and therefore excluded such grains from their study. Empirically, based on the Marhas study and our work, it appears likely that mainstream grains with $\delta^{135}$Ba values above −400 ‰ are contaminated with Solar System barium. Three of the "uncontaminated" grains of Barzyk et al. (2007a) have $\delta^{135}$Ba values above −400 ‰, but all three have large analytical uncertainties (>±160 ‰).

The barium isotopic compositions of Murchison mainstream SiC grains from this study are compared with data from previous studies in Fig. 3 using the selection criterion of $2\sigma(\delta^{135}\text{Ba}) < 160$ ‰ as described above. Eight of the 15 grains measured by Savina et al. (2003a; purple squares in Fig. 3) show $\delta^{135}$Ba above −400 ‰ indicating probable solar system barium contamination. In comparison, all our selected mainstream and unclassified grains except G379 show strongly negative $\delta^{135}$Ba (below −400 ‰), in good agreement with the known uncontaminated grains from previous studies. The unclassified grain, G379, was the largest grain on the mount (3×7 *μm*), and had almost solar barium isotopic composition ($\delta^{135}$Ba = −47±42 ‰). We were not able to determine carbon or silicon isotope ratios for this grain and therefore cannot classify it.

Two of the nine mainstream SiC grains from Indarch meteorite in Fig. 3 have $\delta^{135}$Ba values greater than −400 ‰ (Jennings et al., 2002). Indarch is an enstatite chondrite, which has mineralogical indicators of formation in a reduced environment (Keil 1968). It was argued that Indarch SiC grains are likely to be less contaminated due to the lack of aqueous alteration on the parent body (Barzyk et al. 2007a). However, a recent paper has argued that enstatite chondrites formed under conditions similar to those of other chondrites and were then exposed to a hydrogen-poor, and carbon-, sulfur-rich gaseous reservoirs (Lehner et al. 2013). If this is the case, parent body barium contamination could be similar for presolar SiC grains from both the Murchison and Indarch meteorites. On the other hand, the fact that Murchison grains from Savina et al. (2003a) show much more contamination than the Murchison grains from Barzyk et al. (2007a) indicates laboratory and/or Murchison parent body barium contamination of varying degrees in previous studies. A recent study of barium isotopic composition in mainstream grains by Ávila et al. (2013) reported close-to-solar barium isotopic composition for 12 large SiC grains (7−58 *μm*). Contamination cannot be excluded in their study, especially considering the extremely low barium concentrations in these large grains. Barium contamination in presolar SiC grains is therefore likely caused by both laboratory chemical procedures and alteration/metamorphism on the parent body. Our data indicates that the acid cleaning procedure used in this study effectively removes surface-sited terrestrial and/or parent-body contamination.

### *3.2. Comparison with Previous Data in Single SiC Grains*

Barium isotopic compositions in mainstream SiC grains from this study are less scattered than those from previous studies due to improved stability of laser beams and the instrument. For instance, the new data tend to form a straight line on the $\delta^{137}$Ba versus $\delta^{135}$Ba plot with all values in good agreement with grain aggregate results as shown in Fig. 4a. Implantation of cesium into grains by the NanoSIMS in the Marhas et al. (2007) study resulted in interference with $^{134}Ba^{+}$ due to the formation of $^{133}CsH^{+}$ ions even at high vacuum conditions. If SiC grains were bombarded with a $Cs^{+}$ beam prior to RIMS measurements (e.g., Savina et al. 2003a), the tail of a large $^{133}$Cs secondary ion peak that is not completely suppressed in RIMS extends to mass 134 and causes an interference. Since NanoSIMS analysis was done last in this study, there is no cesium interference in our RIMS spectra.

### *3.3. Torino Postprocessing versus FRUITY AGB Models*

#### *3.3.1. Torino Postprocessing AGB Model*

An in-depth description of the postprocessing AGB model calculations adopted here is given by Gallino et al. (1998). The profile of the main neutron source, $^{13}C(\alpha,n)^{16}O$, i.e., the distribution of $^{13}C$ mass fraction with mass in a one-dimensional model, is poorly constrained by theory. This is caused by several uncertainties affecting AGB stellar models, most importantly, those related to the treatment of convective instabilities. In the current postprocessing model calculations, a $^{13}C$ pocket with a decreasing distribution profile of $^{13}C$ and $^{14}N$ is therefore artificially introduced. A schematic graph of the $^{13}C$ pocket is given in Fig. 1 of Gallino et al. (1998). All the parameters of this $^{13}C$ pocket are listed in Table 2. It is subdivided into three zones (I, II and III) with fixed mass fractions of $^{13}C$, $X(^{13}C)$ and $^{14}N$, $X(^{14}N)$ in each zone to allow the AGB model to reproduce the *s*-process main component in the solar system (A>90) based on the so-called mean neutron exposure ($\tau_0$) (Clayton 1968; Gallino et al. 1998). It was later shown that the nucleosynthesis predictions obtained by using zoned or constant $^{13}C$ profiles provide comparable *s*-process element distributions. For instance, the *s*-process index [hs/ls] obtained for different metallicities (*Z*) is almost the same for zoned and constant $^{13}C$ profiles (Busso et al. 2001). [hs/ls] is the log of the ratio of heavy-*s* (barium-peak) to light-*s* (zirconium-peak) elements, divided by the same ratio in the Solar System.

The $^{13}C$ pocket structure is fixed in the calculations and the parameters of the $^{13}C$ pocket in the standard (ST) case are given in Table 2 for reference. The naming of the ST case derives from the fact that the Solar System *s*-process pattern is best reproduced by averaging 1.5 $M_\odot$ and 3 $M_\odot$ AGB model yields of the ST case at half-solar metallicity (Arlandini et al. 1999). The model calculation starts with the $^{13}C$ pocket in the ST case. The fixed mass fractions of $^{13}C$ and $^{14}N$ in all three zones in the ST case are then multiplied by a factor in order to obtain different mean neutron exposures in D3−U2 AGB model calculations, which are named as $^{13}C$ efficiencies in the literature. In the most recently updated Torino model calculations, this number is linearly scaled from 0.3 to 1.8 for D3 to U2 cases (D3: 0.31; D2: 0.48; D1.5: 0.62; ST: 1.0; U1.3: 1.25; U2: 1.81). Since the total mass of the $^{13}C$ pocket is also fixed in the calculations, the values obtained by multiplying different mass fractions of $^{13}C$ and $^{14}N$ by the constant $^{13}C$ pocket mass therefore simply correspond to different amounts of $^{13}$C and $^{14}$N for the *s*-process nucleosynthesis calculations. We call this model 'Three-zone' hereafter to distinguish it from the model calculations with another $^{13}$C profile discussed in the following sections, in which, only Zone-II contains $^{13}$C. The current postprocessing AGB models have been updated with the most recent cross-section measurements for the entire nucleosynthesis network (see KADoNiS[1]). Recommended solar abundances by Lodders et al. (2009) are adopted for initial input in model calculations. For *r*-mostly isotopes, their initial abundances are higher than solar values (e.g., $^{135}$Ba) in half solar metallicity calculations because of consideration of Galactic Chemical Evolution (GCE) (Bisterzo et al. 2011).

### *3.3.2. FRUITY Model*

In this work, we also consider the AGB nucleosynthesis calculations from the FRUITY database (FRANEC Repository of Updated Isotopic Tables & Yields)[2]. Details of these models are given by Cristallo et al. (2009; 2011). In particular, while the $^{13}$C pocket is introduced in the calculations as a free parameter in the Torino postprocessing models, it self-consistently forms after Third Dredge-Up (TDU) episodes in the FRUITY models (see Cristallo et al. 2009 for details). Cristallo et al. (2011) pointed out that the weighted average $^{13}$C efficiency in FRUITY is comparable to the ST case in the Three-zone postprocessing model for a $Z_{\odot}$, 2 $M_{\odot}$ AGB star. One significant difference between the two AGB model calculations is that the mass of the $^{13}$C in the pocket is constant after each TDU episode in the Torino model, whereas it varies in the FRUITY calculations following the natural shrinking of the helium intershell region. Compared to the KADoNiS database used in the Torino postprocessing calculations, a list of neutron capture rates from Bao et al. (2000) is adopted in FRUITY (see Cristallo et al. 2009 for more detail). For the initial input, recommended solar abundances by Lodders (2003) are adopted in FRUITY.

### *3.4. Barium Isotopic Compositions of Mainstream Grains versus AGB Model Calculations*

In this section, we discuss the effect of stellar masses and metallicities on barium isotope ratios in nucleosynthesis calculations of 1.5 $M_{\odot}$ to 3 $M_{\odot}$ AGB stars with close-to-solar metallicity. The 2 $M_{\odot}$, 0.5 $Z_{\odot}$ AGB Three-zone model is chosen as representative for comparison with mainstream grain data in this study.

---

[1] KADoNis: Karlsruhe Astrophysical Database of Nucleosynthesis in Stars, website http://www.kadonis.org/, version v0.3;

[2] FRUITY database, website http://fruity.oa-teramo.inaf.it/.

*3.4.1. Effects of Mass and Metallicity of AGB Stars on Barium Isotope Ratios*

Previous studies concluded that mainstream SiC grains came from AGB stars of about 1.5−3 $M_{\odot}$ with close-to-solar metallicities (Hoppe et al. 1994; Zinner 2004; Barzyk et al. 2007a). The major effect of AGB star progenitor mass is the increasing contribution from the minor neutron source, $^{22}$Ne($\alpha$,$n$)$^{25}$Mg, with increasing initial mass. This affects the *s*-process isotopic pattern due to more effective activation of neutron-capture channels at various branch points (e.g., $^{134}$Cs as discussed above; see also Käppeler et al. 2011). In Fig. 5, we compare grain data with model predictions of low mass AGB stars with close-to-solar metallicity in three-isotope plots of δ$^{134}$Ba versus δ$^{135}$Ba. The δ$^{134}$Ba values are affected by the $^{134}$Cs branch point and therefore are good indicators of the $^{22}$Ne($\alpha$,$n$)$^{25}$Mg efficiency in the parent AGB stars. In this paper, we plot model predictions for all TPs during the AGB phase as lines starting at the star's assumed initial isotopic composition, but symbols are only plotted for TPs with envelope C/O > 1 for comparison with grain data, since this is when SiC is expected to condense based on thermodynamic equilibrium calculations (Lodders & Fegley 1995).

As discussed in Straniero et al. (2003), the peak temperature within the convective zone powered by a TP depends on core mass and initial metallicity. It is also mildly affected by the erosion of the hydrogen-rich envelope caused by mass loss. For AGB stars with close-to-solar metallicity, the core mass is quite similar in the models with initial masses ranging between 1.5 and 2.5 $M_{\odot}$. Therefore, in such AGB stars the peak temperature within the convective zone powered by a TP is similar (e.g., Straniero et al. 2003). The 2 $M_{\odot}$, 0.5 $Z_{\odot}$ model predictions for δ$^{137}$Ba versus δ$^{135}$Ba (Fig. 4a) are grouped on a nearly straight line, in agreement with the single grain and the aggregate data. The 3 $M_{\odot}$ AGB model with lower-than-solar metallicity needs to be considered separately, since it is characterized by a larger core mass and higher peak temperature at the bottom of the TPs. In Fig. 4b, the upward bending at the tail of the 3 $M_{\odot}$ model calculations for δ$^{137}$Ba values reveals this effect. It is caused by the opening of the branching at $^{136}$Cs at higher stellar temperature (Fig. 1), which results in reduced $^{136}$Ba production and therefore increased δ$^{137}$Ba values compared to 2 $M_{\odot}$ model predictions. Because of their similar core mass, the 1.5 $M_{\odot}$, 0.5 $Z_{\odot}$ model predictions are quite similar to 2 $M_{\odot}$, 0.5 $Z_{\odot}$ ones, as shown in Figs. 5a, c. We will consider the predictions from the 2 $M_{\odot}$ AGB model as representative of low mass AGB stars.

The $^{13}C$ neutron source is primary; the amount of $^{13}C$ depends on the number of protons mixed into the helium intershell that are captured by primary $^{12}C$ generated in the TP by partial helium burning. In general, the higher the metallicity, the lower the core mass and, in turn, the lower the peak temperature during a TP (Straniero et al. 2003). In AGB stars of lower-than-solar metallicity, the convective envelope starts with less oxygen and thus becomes carbon-rich after fewer TPs than a solar metallicity star, as the carbon-rich phase in the 2 $M_{\odot}$, 0.5 $Z_{\odot}$ model (Fig. 5c) is longer (i.e., extends over more TPs) than that in the 2 $M_{\odot}$, $Z_{\odot}$ model (Fig. 5d); this can also be seen in Fig. 6 for FRUITY predictions with different metallicities. We compared the grain data with $Z_{\odot}$ and 0.5 $Z_{\odot}$ model predictions and chose 0.5 $Z_{\odot}$ as representative because the carbon-rich phase of the 0.5 $Z_{\odot}$ model is more extended and better covers the range of the grain data. We therefore compare grain data with the Torino postprocessing model predictions of 2 $M_{\odot}$, 0.5 $Z_{\odot}$ AGB stars, but it does not necessarily mean that all the grains only came from these AGB stars. The model predictions are shown with a range of $^{13}C$ efficiencies from D3 to U2 cases. We assume no contribution to $^{135}Ba$ from decay of $^{135}Cs$, a subject discussed in Section 4.6.

*3.4.2. $\delta^{134}Ba$ versus $\delta^{135}Ba$*

The variation of $\delta^{134}Ba$ values in presolar SiC grains shown in Fig. 5 reflects different neutron exposures experienced in their parent AGB stars. Good agreement between the grain data and the postprocessing calculations is observed. Exceptions are the $\delta^{134}Ba$ values of two mainstream SiC grains that are strongly negative, even outside 2σ uncertainties (red points in Figs. 5 & 6). The δ-values of the other barium isotopes of these two grains are all within the range of the mainstream SiC grains studied. The lowest $\delta^{134}Ba$ value reached in Three-zone calculations is ~ −90 ‰ for the U2 case, but this case predicts positive $\delta^{138}Ba$ values, which are not seen in any grains; the two peculiar grains have $\delta^{138}Ba$ values indistinguishable from other mainstream grains (Fig. 6). In fact, all the AGB model predictions (both Torino and FRUITY) disagree with these two peculiar grains as shown in Figs. 5 & 6. We will discuss why AGB model calculations do not predict strongly negative $\delta^{134}Ba$ values and explore another possible stellar origin in Sections 4.4 and 4.5.

*3.4.3. $\delta^{137}Ba$ versus $\delta^{135}Ba$*

Mainstream SiC grains form a straight line in the plot of $\delta^{137}Ba$ versus $\delta^{135}Ba$ (Fig. 4) while the predictions show an upward bending toward the end of AGB phase due to partial activation of $^{22}Ne$ neutron source at stellar temperatures above $3\times10^{8}$ $K$ (Lugaro et al. 2003a), especially in 3 $M_⊙$, 0.5 $Z_⊙$ AGB stars as shown in Fig. 4b. We calculated the linear regression line of all grain data, including mainstream, AB and Z grain, using Orthogonal Distance Regression (ODR) fit[3] in Igor software with 95 % confidence shown as grey area in Fig. 4a. The linear fitting of single SiC grains is in excellent agreement with that of SiC aggregate data whose uncertainty is negligible (Prombo et al. 1993). Despite differences between samples and techniques, the general agreement points towards a systematic offset of the model predictions. Indeed, the discrepancy could be solved by increasing the $^{137}$Ba cross-section by 30 % (Gallino et al. 1997). Koehler et al. (1998), however, remeasured this cross-section and confirmed the previous value. One alternative is that this discrepancy could result from the present uncertainties in the $^{22}$Ne($\alpha$,$n$)$^{25}$Mg and $^{22}$Ne($\alpha$,$\gamma$)$^{26}$Mg rates, and of their relative efficiency. For instance, the postprocessing model calculations with a $^{22}$Ne($\alpha$,$n$)$^{25}$Mg rate one-half that of the lower limit from Käppeler et al. (1994) predict slightly lower $\delta^{137}$Ba values for the last several TPs and agree better with the grain data, indicating less efficient $^{22}$Ne($\alpha$,$n$)$^{25}$Mg neutron source operating in AGB stars. The discussion in uncertainties in the $^{22}$Ne($\alpha$,$n$)$^{25}$Mg and $^{22}$Ne($\alpha$,$\gamma$)$^{26}$Mg rates and our derived constraints based on $\delta^{134}$Ba values in mainstream grains are given in Section 4.4.

#### *3.4.4. $\delta^{138}$Ba versus $\delta^{135}$Ba*

The neutron-magic isotope $^{138}$Ba ($N$ = 82) acts as a bottleneck in the *s*-process path due to its extremely small neutron-capture cross-section. Its abundance strongly depends on the strength of the major neutron source $^{13}$C($\alpha$,$n$)$^{16}$O during interpulse periods in AGB stars at 8 *keV* (~$10^8$ *K*, Gallino et al. 1998; Lugaro et al., 2003a). In Fig. 7, six out of the 61 grains measured in this study show a strong depletion of $^{138}$Ba (< −400 ‰). As shown in Fig. 3, grains with $\delta^{138}$Ba values below −400 ‰ were found in all previous single-SiC studies except Savina et al. (2003a). In contrast to single grain results, the 'G-component' derived from aggregate measurements is −331 ‰ (Prombo et al. 1993), which is distinctly above −400 ‰. This contradiction can be explained by the scatter of single grains in the three-isotope plot, reflecting a spread of efficiency for the neutron source $^{13}$C in the helium intershell of parent AGB stars. Thus, aggregate studies by measuring isotopic compositions of large quantities of SiC grains cannot resolve variations of $\delta^{138}$Ba values in different individual grains that came from different parent AGB stars.

## 4. DISCUSSION

### *4.1 $\delta^{138}$Ba in Mainstream SiC: A Tracer of $^{13}$C Pocket Structure*

---

[3] The ODR fit is a linear least-squares fit, which considers uncertainties in both x-axis and y-axis for each data point.

In this section, we (1) summarize existing observational constraints on the $^{13}C$ pocket, (2) discuss effects of $^{13}C$ pocket profile and $^{13}C$ pocket mass on model predictions of $\delta^{138}Ba$ and demonstrate the necessity of a smaller $^{13}C$ pocket with a flat $^{13}C$ profile to explain $\delta^{138}Ba <$ −400 ‰ measured in some of the acid-cleaned mainstream grains, and (3) explore possible physical mechanisms that could flatten a $^{13}C$ profile in the pocket.

### *4.1.1. Previous Constraints on the $^{13}C$ Pocket*

Historically, three lines of evidence provided constraints on the range of mean neutron exposures in the $^{13}C$ pocket in AGB stars: the solar system *s*-process pattern, spectroscopic observations of [hs/ls] and [Pb/hs] ratios in stars with different metallicities, and isotopic compositions of heavy elements in presolar grains. Several problems however, are associated with these lines of evidence. (1) Although solar *s*-process abundances are well known, the pattern is the result of nucleosynthesis in all previous generations of AGB stars prior to the formation of solar system 4.56 *Ga* ago. Thus, the distribution of solar *s*-process isotopes is not the signature of a single AGB star. Reproduction of the solar *s*-process pattern therefore requires coupling GCE with AGB stellar model calculations (Travaglio et al. 1999), which may also be affected by uncertainties associated with GCE assumptions. (2) Spectroscopic observations show scatter in the [hs/ls] and [Pb/hs] ratios across AGB stars with comparable metallicities. A range of neutron exposures is required to explain the scatter (Busso et al. 2001; Bisterzo et al. 2010), which could be due to, e.g., different initial masses (e.g., Gallino et al. 1998), or different stellar rotational velocities of AGB stars (Herwig et al. 2003; Siess et al. 2004; Piersanti et al. 2013)[4]. Spectroscopic observations from AGB stars mainly provide elemental abundances; only a few isotopic ratios are available, such as $^{12}C/^{13}C$ and $^{14}N/^{15}N$ (e.g., Hedrosa et al. 2013). Concerning heavy isotopes, Lambert et al. (2002) measured the $f_{odd}$ ([N($^{135}$Ba) + N($^{137}$Ba)] / N(Ba)) value (0.31±0.21) in the star HD 140283 with about a factor of two uncertainty, which is much less precise than presolar grain data. Restrictive constraints therefore cannot be obtained from spectroscopic observations of isotope ratios of barium. (3) Many previous studies of heavy elements in presolar grains were likely affected by contamination. Isotopic measurements of heavy elements in uncontaminated mainstream SiC grains can provided useful constraints on AGB model calculations at the isotopic level (Barzyk et al. 2007a). More recent isotopic ratios measured in clean SiC grains with high precision (see Section 2) provide a unique and much improved tool to constrain AGB model calculations.

[4] The upper limit of [hs/ls] ratios in solar-like AGB stars in the Galactic disk can be explained by enhanced $^{12}C$ abundance in the helium intershell, which increases the maximum $^{13}C$ amount per iron seed in the $^{13}C$ pocket (Lugaro et al. 2003b).

### *4.1.2. How to Reach $\delta^{138}$Ba < −400 ‰ in AGB Model Calculations*

Several branches of the *s*-process path in the cesium-barium region join at $^{136}$Ba, so production of $^{137}$Ba and $^{138}$Ba is little affected by those branchings (Lugaro et al. 2003a). We find that in the Torino model calculations, values of $\delta^{138}$Ba are largely unaffected by uncertainties in the $^{22}$Ne($\alpha$,$n$)$^{25}$Mg rate. Values of barium neutron capture Maxwellian-Averaged Cross-Sections (MACSs), defined as $<\sigma v>/v_T$ (where σ, $v$ and $v_T$ are the neutron capture cross section of a nuclide, the relative neutron velocity, and the mean thermal velocity, respectively), are well determined experimentally with uncertainties between 3 % and 5 % (see KADoNiS for more detail). FRUITY model predictions (Figs. 6) are comparable to Three-zone model predictions (Fig. 7a,b) for $\delta^{138}$Ba. The FRUITY results however are confined to a much smaller range in $\delta^{135}$Ba compared to the Three-zone calculations; variations in masses and metallicities of parent stars in FRUITY calculations are not sufficient to account for the whole range of isotopic compositions of the mainstream grains. This difference might result from different treatments of the $^{13}$C neutron source in the two models as well as difference in the adopted AGB stellar models. The values of $\delta^{138}$Ba below −400 ‰ observed in six grains from our study cannot be reached by the range of $^{13}$C efficiencies in the Torino Three-zone calculations or by the FRUITY predictions. Thus, it indicates a significant source of inaccuracy in the AGB model calculations, which must be explored in order to reach lower $\delta^{138}$Ba values.

In the classical approach for the *s*-process nucleosynthesis calculations, equilibrium in the neutron-capture flow is obtained between magic neutron numbers (e.g., B$^2$FH, 1957). The $<\sigma>N_s$ curve (the MACS of an *s*-only nuclide times its abundance) is therefore approximately constant in these regions. Due to their small MACSs, neutron magic nuclei ($N$ = 50, 82, and 126) behave as bottlenecks in the main *s*-process path, which results in three distinct steps in the $<\sigma>N_s$ curve for *s*-only nuclei in the solar system. As shown in Fig. 2 of Käppeler et al. (2011), a steep decline of $<\sigma>N_s$ values exists between $^{136}$Ba and $^{138}$Ba for the solar *s*-process pattern, which causes the classical gap of [hs/ls] values. This gap results from the fact that *s*-only $^{136}$Ba defines the end point of the *s*-process chain in the region starting at the first *s*-process peak at neutron-magic $^{88}$Sr ($N$ = 50); Similarly, $^{138}$Ba ($N$ = 82) is the starting point of the *s*-process chain between the 2$^{nd}$ and 3rd $s$-process peaks. As shown below, this makes $\delta^{138}$Ba values sensitive to the $^{13}$C profile and the mass of the $^{13}$C pocket.

*4.1.3. Zone-II AGB Model Calculations*

We did postprocessing AGB model calculations with single-zone $^{13}$C pockets based on the Three-zone $^{13}$C pocket in the ST case with the parameters listed in Table 2. The single-zone $^{13}$C pocket contains Zone-II only, with Zone-I and Zone-III being excluded. Zone-II has both relatively higher mass and more $^{13}$C compared to the other two zones and is able to yield enough neutrons for $s$-process nucleosynthesis by itself. Calculations were done with a range of Zone-II $^{13}$C pocket masses ($2.1\times10^{-4}$–$1.0\times10^{-3}$ $M_{\odot}$). Based on our single-zone test results, we found that values of $\delta^{138}$Ba below −400 ‰ are only achievable with a smaller Zone-II $^{13}$C pocket. All the results are shown in Table 3. Zone-II model predictions with D3-to-U2 $^{13}$C efficiencies are shown in Fig. 7c. Model calculations with Zone-II $^{13}$C pockets are shown as filled symbols to distinguish them from Three-zone calculations, which are shown as open symbols. As shown in Fig. 7, the $\delta^{138}$Ba values produced by Zone-II models depend more strongly on the $^{13}$C efficiency and span a wider range of $\delta^{138}$Ba values than do those produced by the Three-zone models. Models with the Zone-II-only $^{13}$C pocket are in better agreement with the grain data, allowing $\delta^{138}$Ba to reach values below −400 ‰.

Abundances of $s$-only $^{136}$Ba and $s$-mostly $^{138}$Ba increase with increasing $^{13}$C amount in the pocket. On the other hand, the variation of $\delta^{138}$Ba values depends on the relative increase of $^{138}$Ba to $^{136}$Ba from D3 to U2 cases, which is affected by both the $^{13}$C profile and the mass of the $^{13}$C pocket. The sensitivity of $\delta^{138}$Ba is caused by its extremely small MACS, which is a factor of ten lower than that of $^{136}$Ba. Thus, the $s$-process equilibrium, $\langle\sigma_A\rangle N_{s(A)} = \langle\sigma_{A-1}\rangle N_{s(A-1)}$, cannot be achieved for the $^{138}$Ba abundance, which therefore strongly depends on the $^{13}$C profile within the $^{13}$C pocket and the pocket mass adopted in the AGB models. The fact that $\delta^{138}$Ba values dropped from the D1.5 (~−100 ‰) to the ST (~−400 ‰) case in Zone-II calculations results from the fact that $^{136}$Ba is enriched by 50 % with respect to $^{138}$Ba ($^{136}$Ba and $^{138}$Ba abundances are increased by factors of 3.36 and 2.26, respectively) from D1.5 to ST, corresponding to a steeper slope for the second step of the $\langle\sigma\rangle N_s$ curve (Fig. 4 of Clayton et al. 1961, Fig. 2 of Käppeler et al. 2011).

Even lower $\delta^{138}$Ba values can be obtained by decreasing the mass of Zone-II by a factor of 2.5, from $5.2\times10^{-4}$ $M_{\odot}$ (Zone-II in Table 3) to $2.1\times10^{-4}$ $M_{\odot}$ (Zone-II_d2.5 in Table 3) as shown in Fig. 7d. As a matter of fact, ST is the only case in which $\delta^{138}$Ba for a reduced mass of Zone-II-only $^{13}$C pocket decreases to more negative values; in the U2 and U1.3 cases, they remain almost the same, and in the D3 to D1.5 cases, the predictions are much closer to the solar system value due to less 'G-component' barium produced by the *s*-process in the helium intershell. Moreover, we did Zone-II calculations with a finer grid of $^{13}$C efficiencies around the ST case to search for lower $\delta^{138}$Ba values and failed. Thus, the effect of reducing Zone-II mass cannot be compensated by varying $^{13}$C efficiencies. All the grains can be well matched by model calculations with the reduced Zone-II mass (defined as Zone-II_d2.5 model in Table 3), except the unclassified grain G260 (Fig. 7). We therefore adopted $2.1\times10^{-4}$ $M_{\odot}$ as the lower limit for Zone-II $^{13}$C pocket mass. Good agreement is maintained for $\delta^{134}$Ba and $\delta^{137}$Ba versus $\delta^{135}$Ba plots. Based on these calculations, a smaller $^{13}$C pocket with a Zone-II-only $^{13}$C profile in the $^{13}$C efficiency range of D3−U1.3 is needed to match barium isotopic composition in all the presolar grains from this study.

As noted in Section 4.1.1, high precision isotopic data from presolar grains provide unique constraints on AGB models. Table 4 gives values of $\tau_0$, [hs/ls], [ls/Fe], [hs/Fe], and [Pb/Fe] for Zone-II model calculations from the U2 to D3. These are comparable to results from the same models using the standard Three-zone $^{13}$C pocket. For instance, values of $\tau_0$ range from 0.15 to 0.51 in the 2 $M_{\odot}$, 0.5 $Z_{\odot}$ Zone-II models shown in Table 4, and from 0.11 to 0.32 in the corresponding Three-zone models (not shown). The values of spectroscopic observables such as [hs/ls] are also largely unaffected by the $^{13}$C profile within the $^{13}$C pocket and the pocket mass; [hs/ls] values range from −0.50 to 0.62 in Zone-II models, and −0.55 to 0.35 in Three-zone models.

We also did postprocessing calculations with smaller Three-zone $^{13}$C pockets; the results in the ST case are shown in Table 2. Similar to Zone-II models, the Three-zone models predict that $\delta^{138}$Ba values decrease with decreasing mass of the Three-zone $^{13}$C pocket down to Three-zone_d2.5 model shown in Fig. 7b, below which $\delta^{138}$Ba barely changes. Even so, values of $\delta^{138}$Ba are all above −400 ‰ in the Three-zone calculations. Due to the relative uncertainties in the neutron capture MACS values for $^{136}$Ba and $^{138}$Ba (≤ 5 %), the uncertainty in $\delta^{138}$Ba predictions is ±50 ‰ (2σ uncertainty) at most. Therefore, based on the tests done so far, not only a smaller, but also a flat (Zone II-like) $^{13}$C pocket is required to explain the grains with $\delta^{138}$Ba values below −400 ‰. To summarize, Three-zone to Three-zone_d2.5 models are able to match the majority of the grain data with $\delta^{138}$Ba values above −400 ‰. On the other hand, Zone-II to Zone-II_d2.5 models can match all the grain data for $\delta^{138}$Ba. Since it is highly likely that different $^{13}$C pockets exist in the parent AGB stars of mainstream grains, we point out the fact that the Zone-II $^{13}$C pockets are only required to explain the grains with $\delta^{138}$Ba < −400 ‰.

#### *4.1.4. Flattening the Distribution of $^{13}$C in the $^{13}$C Pocket of AGB Stars*

In our calculations a smaller $^{13}$C pocket with a flat $^{13}$C abundance profile provides better agreement with the grain data. There are several possible mixing mechanisms that could yield flat $^{13}$C profiles. Rotation-induced mixing may lead to a partial mixing of $^{13}$C and $^{14}$N within a $^{13}$C pocket (Herwig et al. 2003; Siess et al. 2004; Piersanti et al. 2013 and references therein). In particular, meridional circulation (Eddington-Swift instability, ES) may smooth and enlarge the tail of the newly formed $^{13}$C pocket, where the $^{13}$C mass fraction is larger than that of $^{14}$N. Note that the tail of the newly formed $^{13}$C pocket is the inner region of the pocket, where most of the *s*-process nucleosynthesis takes place (see Fig. 2 of Piersanti et al. 2013 for details). In principle, this occurrence could justify the assumption of a nearly flat $^{13}$C profile. On the other hand, the presence of a Goldreich-Schubert-Fricke (GSF) instability in the top layers of the $^{13}$C pocket (towards the convective envelope) may induce an inward mixing of $^{14}$N, which is a major neutron poison, and thus reduces the number of neutrons available for the synthesis of heavy elements. Note that in this case it is not the $^{13}$C profile itself that is flattened, but rather resulting neutron density. Since stars with similar initial mass and metallicity may have different initial rotational velocities, the effect of rotation on *s*-process nucleosynthesis is expected to vary from star to star, giving rise, among the other possibilities, to a spread in the barium isotopic composition of the envelopes of those stars. Piersanti et al. (2013) studied AGB stars of metallicity Z = 0.014 and Z = 0.0007 with different initial rotational velocities and found that when a moderate rotational velocity is assumed at the beginning of the main sequence, the resulting meridional circulation may produce modifications of the *s*-process during the AGB phase. This is true for stars with low metallicity (Z = 0.0007) only, while most of the mainstream presolar grains presumably originate in close-to-solar metallicity (Z = 0.014) carbon stars. On the contrary, the GSF instability is active at any metallicity, and moves part of $^{14}$N from the more external portion of the pocket down to the tail where $^{13}$C is higher than $^{14}$N and most of the *s*-process takes place (see Fig. 2 of Piersanti et al. 2013 for details). Piersanti et al. (2013) found that the main effect of a moderate rotation in AGB models with nearly solar composition is an increase of $\delta^{135}$Ba, while the other isotope ratios, $\delta^{134}$Ba in particular, are almost unchanged. A significant reduction in $\delta^{138}$Ba is found in rotating AGB model calculations with $Z = 0.01$: −350 ‰ for a rotational speed of 30 km s$^{-1}$ versus −200 ‰ in the nonrotational case. While these calculations provide an indication that rotation may have an important influence, at present it seems not to be the mechanism responsible for $\delta^{138}$Ba values below −400 ‰, which is observed in ~10 % of the mainstream grains in this study and previous studies.

The existence of a nonnegligible magnetic field could potentially affect the shape and size of the $^{13}$C pocket. Indeed, magnetic buoyancy has been proposed as an alternative mechanism for forming the $^{13}$C pocket in low-mass AGB stars (Busso et al. 2012). Since this mechanism may continue to operate during the interpulse period, the resulting mixing might produce larger and more flattened $^{13}$C pockets. Such a possibility deserves further investigation.

### *4.2. Effects of Flatter $^{13}$C Pockets on Some Other s-Process Isotopes*

We compared Zone-II and Three-zone model predictions in a 2 $M_{\odot}$, 0.5 $Z_{\odot}$ AGB star for isotopes of other elements, and found that $\delta$($^{88}$Sr/$^{86}$Sr), $\delta$($^{138}$Ba/$^{136}$Ba) and $\delta$($^{208}$Pb/$^{206}$Pb) at the three *s*-process peaks are extremely sensitive to the $^{13}$C pocket profile and the pocket mass. The sensitivity is caused by the extremely small MACS values at 30 *keV* of neutron-magic $^{88}$Sr (6.13±0.11 *mb*), $^{138}$Ba (4.00±0.20 *mb*) and $^{208}$Pb (0.36±0.03 *mb*). By comparing Zone-II model predictions to previous presolar grain data for other elements, we observe that better agreement is obtained for zirconium isotopes, $\delta$($^{92}$Zr/$^{94}$Zr) in particular. A detailed comparison of the grain data with the Zone-II models for zirconium isotopes will be given elsewhere. Good agreement remains for molybdenum and ruthenium isotopes and for $\delta$($^{87}$Sr/$^{86}$Sr) (Nicolussi et al. 1997, 1998; Savina et al. 2004; Barzyk et al. 2007a). The U2 case for Zone-II calculations yields $\delta$($^{90}$Zr/$^{94}$Zr) values that are too negative and $\delta^{138}$Ba values that are too positive to match the grain data (see Fig. 7 for Ba), so this case can therefore be safely excluded from our discussion.

Large enhancements of $\delta$($^{88}$Sr/$^{86}$Sr) values are found in Zone-II calculations. The ST and U1.3 cases in the Zone-II calculations predict $\delta^{88}$Sr values as high as 1400 ‰ and 2500 ‰, respectively. In contrast, the highest $\delta^{88}$Sr values measured in mainstream SiC grains (281±118 ‰) are much lower than the predictions with the lower limit of the $^{22}$Ne($\alpha$,$n$)$^{25}$Mg rate from Käppeler et al. (1994), obtained after excluding the elusive 635 keV resonance contribution (K94 rate hereafter). The fact that most grains from Nicolussi et al. (1998) lie significantly closer to solar values in $\delta$($^{84}$Sr/$^{86}$Sr) than any of the model predictions is an indicator that those grains are probably highly contaminated. In this work we have solved previous problems with barium contamination in SiC grains. Isotopic measurement of grains with minimal strontium contamination would provide another fundamental contribution to constrain *s*-process nucleosynthesis in AGB stars.

It is important to point out that model predictions have large uncertainties for strontium isotope ratios due to MACS uncertainties of several isotopes in this region. Only theoretical estimates exist for $^{85}$Kr (terrestrial $t_{1/2}$=10.8 *years*), and the most recent predictions disagree with each other by more than a factor of two (Rauscher et al. 2000; Goriely & Siess 2005). For $^{86,87}$Sr, quite old experimental reaction rates are available by TOF and activation measurements (see KADoNiS; Dillmann et al. 2006). In addition, the relative abundances of strontium isotopes are strongly affected by the branch point at $^{85}$Kr; both $^{86}$Sr and $^{87}$Sr increase with a decreasing $^{22}$Ne($\alpha$,$n$)$^{25}$Mg rate, while $^{88}$Sr is unaffected. Changing the $^{22}$Ne($\alpha$,$n$)$^{25}$Mg rate from K94 to ½×K94 causes $\delta^{88}$Sr values for the last TP to drop from 2500 ‰ to 2000 ‰ for the U1.3 case, and from 1400 ‰ to 1050 ‰ for the ST case. We plan to measure correlated strontium and barium isotope abundances in acid-cleaned mainstream SiC grains to better constrain the nuclear and the $^{13}$C pocket uncertainties in the future.

### *4.3. Extra-Mixing Processes during Red Giant Branch (RGB) & AGB Phases*

The $^{12}$C/$^{13}$C ratios range from 30 to 97 ($^{12}$C/$^{13}$C$_{\odot}$ = 89) in the 24 mainstream SiC grains from this study, and are plotted versus $\delta^{135}$Ba in Fig. 8. $\delta^{135}$Ba is chosen because it is relatively independent of model parameters and nuclear input uncertainties (e.g., uncertainty in the $^{22}$Ne($\alpha$,$n$)$^{25}$Mg rate) in nucleosynthesis calculations. The grain data are consistent with both model calculations (Torino postprocessing and FRUITY) except two grains with $^{12}$C/$^{13}$C ~30 that cannot be matched by the model predictions during carbon-rich AGB pulses. In addition, no mainstream SiC grains with $^{12}$C/$^{13}$C >100 are found in this study, although both models predict that 50 % of the stellar mass is lost when $^{12}$C/$^{13}$C >100.

Mixing in AGB stellar models currently lacks an appropriate treatment along boundaries of convective regions. Of particular interest is the nonconvective mixing (extra-mixing) at the base of the convective envelope during the RGB and, possibly, the AGB phase. This mechanism should be able to mix hydrogen-burning processed material ($^{13}$C-rich and, eventually, $^{14}$N-rich) with the convective envelope. When this material is transported back to the surface, lower surface $^{12}$C/$^{13}$C values are attained. An extra-mixing process during the RGB phase is considered in Torino models (initial $^{12}C/^{13}C = 12$), but not in FRUITY models (initial $^{12}C/^{13}C \sim 23$). Neither of the two models considers an extra-mixing process during the AGB phase, such as cool bottom processing (CBP) (Nollett et al. 2003 and references therein). The RGB extra-mixing process is required to attain a $^{12}C/^{13}C$ ratio between 30 and 60 during the AGB phase. Lower values can be attained if extra-mixing is also at work during the AGB phase. Note, however, that around 10 % of mainstream SiC grains studied so far have $10 < ^{12}C/^{13}C < 30$ (WUSTL Presolar Database[5], Hynes & Gyngard 2009). These values could be attained by hypothesizing the activation of CBP, or alternatively, the occurrence of proton ingestion episodes at the end of the AGB phase (see Section 4.5). By introducing the RGB extra-mixing process in the FRUITY calculations, the $^{12}C/^{13}C$ ratios in 2 $M_{\odot}$, 0.7 $Z_{\odot}$ calculations, for instance, would be reduced by roughly a factor of two, in better agreement with the grain data. In addition, only 0.4 % of mainstream SiC grains have $^{12}C/^{13}C > 100$ (WUSTL Presolar Database), while the Torino models yield larger values ($^{12}C/^{13}C \sim 150$) for close-to-solar metallicity AGB stars. However, if lower initial carbon isotope ratios are adopted ($^{12}C/^{13}C = 8$, still compatible with observations, e.g., Fig. 5a of Palmerini et al. 2011), final $^{12}C/^{13}C$ ratios would only reach 100 instead of 150, and would be in agreement with the grain data. Finally, different conclusions can be drawn for presolar oxide grains (e.g., Palmerini et al. 2011), in particular Group 2 corundum grains (~25 % of all presolar corundum, Nittler et al. 1997), which require the activation of extra-mixing during both the RGB and AGB phases. The different percentages of presolar carbon (~10 %) and oxide grains (~25 %) apparently requiring CBP during the AGB phase might result from differences in the initial masses of their parent stars (1.5−3 $M_{\odot}$ for SiC; 1−1.5 $M_{\odot}$ for corundum).

### *4.4. Constraints on the $^{22}Ne(\alpha,n)^{25}Mg$ Rate From $\delta^{134}Ba$ Values in Mainstream Grains*

In this section we consider uncertainties in the nuclear inputs at the $^{134}Cs$ branch point and the $^{22}Ne(\alpha,n)^{25}Mg$ rate in order to evaluate $\delta^{134}Ba$ values in SiC grains. In addition, we discuss the difficulty of predicting negative $\delta^{134}Ba$ at the present stage based on comparison between the grain data, and the Torino and FRUITY model calculations.

#### *4.4.1. Uncertainties of $^{134}Cs$ Neutron Capture and $\beta^-$ Decay Rates*

The MACSs of $^{134}Ba$ and $^{136}Ba$ are well determined experimentally with 6 % and 3 % uncertainties, respectively (see KADoNiS). Therefore, the nuclear uncertainties affecting reproduction of solar main *s*-process $^{134}Ba/^{136}Ba$ ratio in AGB model calculations mainly result from uncertainties in the MACS and $\beta^-$ decay rate of $^{134}$Cs, and in the efficiency of the $^{22}$Ne neutron source. For the $^{134}$Cs MACS, good agreement has been achieved recently (Patronis et al. 2004 (*experiment*): 724±65 *mb*; Goriely 2005 (*theory*): 805 *mb*), although discrepancies existed previously from different theoretical calculations and experiments (Harris 1981; Shibata et al. 2002; Koning et al. 2005). In both Torino postprocessing and FRUITY model calculations, the MACS values from Patronis et al. (2004) are adopted.

[5] WUSTL presolar database, website http://presolar.wustl.edu/~pgd/.

As noted above, the $\beta^-$ decay rate of $^{134}$Cs is highly sensitive to stellar temperature, increasing by a factor of ~65 as the temperature rises from $1\times10^8$ *K* to $3\times10^8$ *K* during TPs (TY87 Table). The $\beta^-$ decay rate of $^{134}$Cs adopted in both AGB model calculations is based on the TY87 table: in FRUITY, a set of $\beta^-$ decay rates derived from the TY87 table is adopted in the temperature range of the convective zone during TPs (~$1\times10^7$–$3\times10^8$ *K* in 2 $M_\odot$ AGB stars); in the Torino model, the $\beta^-$ decay rate is the geometric average of the TY87 table during the advanced convective TPs (by averaging four data points of the TY87 table at $T$ = 0.5, 1, 2, $3\times10^8$ *K* when $^{22}$Ne($\alpha$,$n$)$^{25}$Mg is marginally activated), which is $1.6\times10^{-7}$ $s^{-1}$ for a 2 $M_\odot$, 0.5 $Z_\odot$ AGB star. Considering the uncertainty of the TY87 table estimated by Goriely (1999), the lower and upper limits of the averaged $\beta^-$ decay rate are $7.0\times10^{-8}$ $s^{-1}$ and $3.0\times10^{-7}$ $s^{-1}$, respectively. The $^{134}$Ba/$^{136}$Ba ratio in the solar main *s*-process component can be well reproduced within 3 % when a value of $1.6\times10^{-7}$ $s^{-1}$ is used in the Torino calculations (e.g., Bisterzo et al. 2011).

*4.4.2. Effects of $^{22}$Ne(α,n)$^{25}$Mg Rates on $\delta^{134}$Ba Values*

The experimentally determined reaction rate of $^{22}$Ne($\alpha$,$n$)$^{25}$Mg has large uncertainties at low energy due to the possibility of low-energy resonances (Wiescher et al. 2012). Different recommended $^{22}$Ne($\alpha$,$n$)$^{25}$Mg rates with different orders of magnitude uncertainties have been reported in the literature ( Caughlan & Fowler 1988: 1.86; Käppeler et al. 1994: 9.09; Angulo et al. 1999: 4.06; Jaeger et al. 2001: 2.69; Longland et al. 2012: 3.36; for $T = 3\times10^8$ *K* and a rate unit of $\times10^{-11}$ $cm^3$ $mol^{-1}$ $s^{-1}$). For instance, the uncertainty reported by Angulo et al. (1999) is up to 60 (in the same units) for the rate at $T = 3\times10^8$ *K*. Note that $^{22}$Ne($\alpha$,$n$)$^{25}$Mg is marginally activated around $3\times10^8$ *K*, which corresponds to the maximum temperature reached at the base of the convective shell generated by TPs in 2 $M_\odot$ AGB stars (Bisterzo et al. 2011). On the other hand, the values of the $^{22}$Ne($\alpha$,$\gamma$)$^{26}$Mg rate reported in the literature are in good agreement with each other (e.g., Käppeler et al. 1994: 1.22; Longland et al. 2012: 1.13, for $T = 3\times10^8$ *K* and a rate unit: $\times 10^{-11}$ $cm^3$ $mol^{-1}$ $s^{-1}$). Stringent constraints on the $^{22}$Ne($\alpha$,$n$)$^{25}$Mg rate are needed for AGB model calculations to yield accurate predictions.

Because the MACS values of $^{134}$Ba and $^{136}$Ba have small uncertainties, discrepancies between the measured and calculated $^{134}$Ba/$^{136}$Ba ratio in grains are likely due to uncertainties in modeling the branch point at $^{134}$Cs. If the $^{22}$Ne($\alpha$,$n$)$^{25}$Mg neutron source was not active, *s*-only $^{134}$Ba would be overproduced by 43 % while its *s*-only twin $^{136}$Ba only by 2 %, with respect to the *s*-only isotope $^{150}$Sm (by averaging yields of 1.5 $M_\odot$ and 3 $M_\odot$ AGB stars at half-solar metallicity). As the abundance of $^{134}$Ba is strongly affected by the $^{134}$Cs branch point and sensitive to the activation of the $^{22}$Ne neutron source with that of $^{136}$Ba mainly affected by the $^{13}$C neutron source, $\delta^{134}$Ba values in presolar SiC grains therefore can be used to derive stringent constraints on the uncertainties affecting the branch point, i.e., the uncertainty in the $^{22}$Ne($\alpha$,$n$)$^{25}$Mg rate.

We compare grain data to Torino model calculations with a varying $^{22}$Ne($\alpha$,$n$)$^{25}$Mg rate in Fig. 9. The dominant effect of increasing the rate is to shift $\delta^{134}$Ba values from strongly positive for the ¼×K94 rate (a range of ~200 to 400 ‰) to near solar for the 4×K94 (4×K94 is close to the upper limit from Käppeler et al. 1994). The ¼×K94 case in Fig. 9 can be safely excluded as it misses a whole group of grains with $\delta^{134}$Ba values close to solar; the 2×K94 & 4×K94 cases disagree with grains with relatively higher $\delta^{134}$Ba values and therefore are worse than ½×K94 and K94 cases. Thus, calculations with the $^{22}$Ne($\alpha$,$n$)$^{25}$Mg rate ranging from ½×K94 to K94 agree the best with the grain data.

We also consider the upper and lower limits of $^{134}$Cs $\beta^-$ decay rate and their effects on constraining the $^{22}$Ne($\alpha$,$n$)$^{25}$Mg rate (hereafter model predictions with the upper limit of the $^{134}$Cs $\beta^-$ decay rate are referred to as UL; ones with the lower limit, LL). When the decay rate is increased from our standard choice ($1.6\times 10^{-7}$ $s^{-1}$) to its upper limit ($3.0\times 10^{-7}$ $s^{-1}$), ¼×K94 UL predictions shift to more positive $\delta^{134}$Ba values and the agreement with grain data is worse. The 2×K94 UL predictions are essentially the same as the 2×K94 case in Fig. 9 (less than 50 ‰ increase in $\delta^{134}$Ba values for the last TP while the highest $\delta^{134}$Ba value stays almost the same in both ST and D1.5 cases for 2×K94 UL). Thus, increasing the $^{134}$Cs $\beta^-$ decay rate to its upper limit does not change our constraints on the $^{22}$Ne($\alpha$,$n$)$^{25}$Mg rate. In contrast, when the decay rate is decreased to its lower limit, ¼×K94 and ½×K94 LL predictions are quite similar to ½×K94 and K94 cases, respectively, in Fig. 9; the highest $\delta^{134}$Ba value in 2×K94 LL predictions is about the same as 2×K94 case ($\delta^{134}$Ba is shifted to around −100 ‰ for the last TP). In summary, decreasing the $^{134}$Cs $\beta^-$ decay rate to its lower limit shifts our constraints to ¼×K94−½×K94 rates. Due to the uncertainty in the $^{134}$Cs $\beta^-$ decay rate resulting from the existence of many low-lying states with unknown log *ft* decay strengths (Goriely 1999), a more conservative constraint on the $^{22}$Ne($\alpha$,$n$)$^{25}$Mg rate is from ¼×K94 to K94. As the K94 rate is 4.14×10$^{-11}$ $cm^3$ $mol^{-1}$ $s^{-1}$ at 3×10$^8$ $K$, the median rate corresponds to 2.07×10$^{-11}$ $cm^3$ $mol^{-1}$ $s^{-1}$ with an uncertainty of a factor of two, which is in agreement with the experimental determination by Jaeger et al. (2001) and the recent evaluation by Longland et al. (2012). In addition, the grain data in the plot of $\delta^{137}$Ba versus $\delta^{135}$Ba in Fig. 4 also suggests a lower than the K94 rate as discussed earlier, in good agreement with the constraints from the $\delta^{134}$Ba values in grains.

The lowest $\delta^{134}$Ba value that can be achieved in Zone-II model calculations is around −100 ‰. In FRUITY model calculations, the set of $^{22}$Ne($\alpha$,$n$)$^{25}$Mg rates from Jaeger et al. (2001) is adopted. The rate reported by Jaeger et al. (2001) is a factor of two lower than the K94 rate at $T$ = 3×10$^8$ $K$. As can be seen in Fig. 6, FRUITY predicts $\delta^{134}$Ba values all above zero for C>O, as does the Torino AGB model. FRUITY however only reaches grains with the highest $\delta^{134}$Ba values and agrees poorly with the vast majority of the grains. A distinctive feature of FRUITY model predictions is that negative $\delta^{134}$Ba values are predicted for the convective envelope composition after the first TDU in 2 $M_\odot$ stars with lower-than-solar metallicities. This results from the fact that the major neutron source $^{13}$C is not completely burned during radiative conditions due to the relatively low stellar temperature at this stage; later on, the leftover $^{13}$C burns convectively during the TP phase at higher temperature, which leads to a higher neutron density and therefore a negative $\delta^{134}$Ba value as neutron capture is favored over $^{134}$Cs decay. Predicted negative $\delta^{134}$Ba values in FRUITY disappear when the metallicity increases to 1.5 $Z_\odot$ for 2 $M_\odot$ because TDU efficiency decreases with increasing metallicity; less of the processed material in the helium intershell is brought up to mix with the convective envelope. In 3 $M_\odot$ FRUITY calculations with close-to-solar metallicities, the absence of this phenomenon is due to the effective radiative burning of $^{13}$C during the interpulse phase because of increased stellar temperature at higher stellar mass. To conclude, equal production or overproduction of $^{134}$Ba compared to $^{136}$Ba is a signature of *s*-process nucleosynthesis in AGB stars, and two nominally mainstream grains identified in this study that lie far below the predictions (G244 and G232 in Table 1 and Figs. 6 & 9) likely do not come from AGB stars. Their possible stellar origin is discussed in the next section.

### *4.5. Negative $\delta^{134}$Ba Values: Signature of the i-Process in Post-AGB Stars*

The intermediate neutron capture process (*i*-process) was first introduced by Cowan & Rose (1977) for evolved red giant stars. The typical neutron densities in the *i*-process are $10^{15}$–$10^{16}$ *neutrons cm*$^{-3}$ with the main neutron source being $^{13}C(\alpha,n)^{16}O$, where $^{13}C$ is formed by the ingestion of hydrogen in helium-burning conditions. Rapid burning of ingested hydrogen causes neutron fluxes higher than typical *s*-process densities (~$10^{7}$–$10^{8}$ *neutrons cm*$^{-3}$), but much less than required for the *r*-process (~>$10^{20}$ *neutrons cm*$^{-3}$, e.g., Thielemann et al. 2011 and references therein).

After the loss of their envelope during the AGB phase, remnant stars continue their evolution along the post-AGB track (e.g., Werner & Herwig 2006). The chemical and physical conditions of their progenitor AGB stars are therefore recorded in their surface and in the helium intershell. Based on simulation results, Iben (1984) concluded that ~10 % of the stars leaving the AGB stage undergo a VLTP and become born-again AGB stars during their post-AGB evolution. Sakurai's object is one of the two observed objects that have undergone a VLTP with $^{12}C/^{13}C \approx 4\pm1$ on the surface (Pavlenko et al. 2004). The abundances of 28 elements of Sakurai's object have been determined by Asplund et al. (1999), and show an enhancement at the first *s*-process peak. With the guidance of the observational data, Herwig et al. (2011) investigated hydrogen ingestion nucleosynthesis during a VLTP event for Sakurai's object. In brief, a small amount of hydrogen remaining on the surface of an AGB star is convectively mixed into the helium-burning zone underneath to form a $^{13}C$ neutron source during the VLTP. A neutron density around $10^{15}$–$10^{16}$ *cm*$^{-3}$ is generated, which is significantly higher than that in an AGB $^{13}C$ pocket, and the resulting nucleosynthesis gives rise to an elemental abundance pattern that matches that of Sakurai's object.

Figure 10 shows the result for barium isotopes, along with grains G244 and G232. The model assumes that there is no significant *s*-process contribution during the AGB phase, in agreement with the observation of Sakurai's object. One-dimensional hydrostatic models predict that the helium intershell immediately splits into two different zones by the energy generated by the hydrogen ingested. In contrast, Herwig et al. (2011) assumed a delay of few hours for the occurrence of the split after the ingestion event, allowing the *i*-process to be activated to reproduce the observed anomalous abundances. The split time is therefore considered as a free parameter in the model. Two cases reproducing the general trend of Sakurai's object observations are shown in Fig. 10, with delay times of 800 and 1200 *minutes* respectively.

The model was developed to explain Sakurai's object with many specific assumptions regarding this star, which is not the parent star of these two grains. The discussion below is therefore only qualitative. Uncertainties in the reaction rates affecting the neutron production are also considered and presented as two cases with the split time set at 1000 *minutes*. The weighted average barium isotopic compositions of the helium-burning zone start at solar composition in the three-isotope plots at $t = 0$ and evolve to negative values with time. Unstable cesium isotopes are produced in great abundance during the rapid burning process. Assuming that grains were formed over a period of a few years, we consider that the shorter-lived isotopes of cesium, $^{136}$Cs and $^{138}$Cs (half-lives shown in Fig. 1), decay to barium prior to grain formation. The half-life of $^{134}$Cs is reduced from the terrestrial value of 2.1 *years* to 3.8 *days* at $3\times10^8$ *K*. (The half-life of $^{134}$Cs does not increase significantly as the temperature rises above $3\times10^8$ *K*.) As the hydrogen-ingested nucleosynthesis during the VLTP lasts for around 2 *days* at most according to the model calculations, $^{134}$Cs will not completely decay to $^{134}$Ba at stellar temperature in 2 *days* nor during grain condensation (lower temperature, half-life ~ 2.1 *years*) in a few years. Model predictions with complete $^{134}$Cs decay and without $^{134}$Cs decay are therefore shown in Figure 10 for comparison with the grain data. Figure 10 shows that the trajectories of the model predictions for $^{134}$Ba, $^{135}$Ba, and $^{137}$Ba relative to $^{136}$Ba ratios are comparatively insensitive to variations in split time and reaction rates. The final $^{138}$Ba/$^{136}$Ba ratio, however, is strongly affected. In all cases, the barium isotopic compositions of the two grains with strongly negative $\delta^{134}$Ba values lie on a mixing line (the thin dashed line in Fig. 10) between solar and the final composition of the helium intershell. Thus, nucleosynthesis during VLTP post-AGB phase can produce negative $\delta^{134}$Ba values, in contrast to the positive $\delta^{134}$Ba values produced by the *s*-process nucleosynthesis in AGB stars. It is noteworthy that the large increase in $\delta^{138}$Ba (off-scale in Fig. 10) occurs when $^{138}$Cs is accumulated in high neutron density environments. $\delta^{138}$Ba drops back to negative values when $^{138}$Cs is bypassed in the neutron flow.

A range of $^{12}$C/$^{13}$C ratios is predicted in the VLTP model calculations, which could explain the carbon isotopic compositions of the two grains as shown in Fig. 11 (model predictions of $\delta^{134}$Ba are shown as the case of complete $^{134}$Cs decay in Fig. 10). The helium-burning zone starts with pure $^{12}C$ (the weighted average $^{12}C/^{13}C$ ratio of the zone at $t = 0$ is $10^8$). The $^{12}C/^{13}C$ ratio quickly drops to solar value (~700 *min* after $t = 0$) before the zone-split in the model calculations. The final $^{12}C/^{13}C$ ratio of the helium-burning zone is also affected by the reaction rates of $^{13}C(\alpha,n)^{16}O$ and $^{14}N(n,p)^{14}C$ as shown in Fig. 11, $2\times{}^{13}C(\alpha,n)^{16}O$ and $2\times{}^{14}N(n,p)^{14}C$, respectively, but it mainly depends on the hydrodynamic properties of the hydrogen ingestion event.

The model predicts a large overproduction of $^{30}Si$ ($\delta^{30}Si$ up to 20,000 ‰) during the post-AGB phase, while the silicon isotopic compositions of both grains are within ±100 ‰. The contradiction may result from several factors. (1) Of course, grains G244 and G232 did not come from Sakurai's object itself. The parent star(s) of these grains may have had different initial abundances in the helium-burning zone when hydrogen ingestion starts, as well as different masses and metallicities of AGB precursors. The good agreement with VLTP post-AGB calculations for barium isotope ratios of these two grains indicates nucleosynthesis in higher neutron density than *s*-process stellar environment. Specifically, in post-AGB stars experiencing a Late Thermal Pulse (LTP), neutron-capture nucleosynthesis processes do not occur in the LTP. In such stars, the LTP simply brings almost pure helium intershell material with its *s*-process signature to the surface. Even for the most up-to-date models, it is still unclear if there are intermediate objects between VLTP and LTP with an intermediate neutron density stellar environment that might yield the isotopic signatures of both silicon and barium we observed in these two grains; (2) The model used here is a modified one-dimensional model. Although it succeeded in matching the surface abundances of Sakurai's object, more results from complete multidimensional hydrodynamic simulations (e.g., Woodward et al. 2013) are needed for a quantitative comparison with the products of hydrogen ingestion events. The multidimensional results will also allow us to perform a detailed comparison with presolar grain isotopic compositions of both light and heavy elements measured in individual grains. In summary, the strongly negative $\delta^{134}Ba$ cannot be explained by the current *s*-process calculations in AGB stars. Instead, it is consistent with the *i*-process signature from a post-AGB VLTP event. Comparisons with more post-AGB models based on new hydrodynamic simulations need to be done in the future.

### *4.6. Barium in Mainstream SiC Grains: Condensation or Implantation?*

The unstable nuclide $^{135}$Cs, sitting along the main *s*-process path, decays to $^{135}$Ba with a half-life of 2 *Ma*. Cesium is such a volatile element that it remains in the gas phase when SiC grains condense (Lodders & Fegley 1995). However, cesium could be trapped in SiC grains if implantation occurred after grain formation. Thus, by evaluating $^{135}$Ba abundances in mainstream grains, $^{135}$Cs condensation can be distinguished from implantation in SiC grains. As pure *p*-process nuclei, $^{130}$Ba and $^{132}$Ba are completely destroyed in the *s*-process. Their abundances are therefore dominated by the initial 'N-component' present in the convective envelope of AGB stars and are insensitive to the magnitudes of $^{13}$C and $^{22}$Ne neutron fluxes in AGB stars. The Torino model predicts strongly negative $\delta^{130,132}$Ba values because of $^{136}$Ba overproduction in the *s*-process. The model predictions are straight lines in plots of $\delta^{130,132}$Ba versus $\delta^{135}$Ba, and the slopes are the same within a few ‰ due to the similarity of their initial abundances. We therefore sum the counts of $^{130}$Ba and $^{132}$Ba to increase the signal-to-noise ratios of these low abundance isotopes and to examine if there is radiogenic $^{135}$Ba from $^{135}$Cs decay in the grains.

Lugaro et al. (2003a) concluded that mainstream SiC grains condensed around AGB stars, where most of $^{135}$Cs had not yet decayed to $^{135}$Ba, based on Murchison SiC aggregate data (Prombo et al. 1993). Compared to the aggregate measurements, a single grain study allows unambiguous determination of each grain and provides information about its formation history. We report the $\delta^{130+132}$Ba values of seven single SiC grains in Table 1. Five grains have no carbon or silicon isotope measurement results due to complete consumption during RIMS analysis. Barium isotopic compositions of all seven grains show *s*-process signatures and are all well within the range of barium isotope ratios of other mainstream SiC grains. It is therefore safe to consider these grains as mainstream for the following discussion. In Fig. 12, the seven single grains are compared with the aggregates and model calculations in two extreme cases: complete $^{135}$Cs decay and no $^{135}$Cs decay. The seven grains with grain sizes between 1 and 3 $\mu m$ are well matched to model predictions with no $^{135}$Cs decay within the uncertainties, in good agreement with the aggregate results. To conclude, single mainstream SiC grains likely formed prior to significant $^{135}$Cs decay in the AGB stellar envelope. The absence of radiogenic $^{135}$Ba in single SiC grains contradicts the implantation scenario for barium (Verchovsky et al. 2003, 2004). Implantation efficiency depends on ionic radius and ionization potential. Barium and cesium have similar ionic radii and more importantly, the ionization potential of cesium is lower than that of barium, which makes cesium easier to be ionized and implanted into SiC grains compared to barium. Unless the implantation model is able to decouple barium from cesium, it is likely that barium condensed into mainstream SiC grains in the AGB circumstellar envelope.

## 5. CONCLUSIONS

1. Comparison of barium isotopic compositions of acid-cleaned mainstream grains in this study with previous results from unwashed samples demonstrates that acid washing is effective at removing barium contamination. Chemical extraction of SiC grains in laboratories and/or aqueous alteration/metamorphism on parent bodies are possible contamination sources.
2. $\delta^{138}$Ba values in mainstream SiC grains are a sensitive tracer of the distribution and mass of $^{13}$C in the $^{13}$C pocket and suggest a flatter distribution of $^{13}$C in the pocket than is currently used in AGB models. Systematic offsets of $\delta^{138}$Ba between the grain data and Three-zone AGB postprocessing/ FRUITY model predictions were confirmed in this study. Torino postprocessing AGB model calculations show that although predictions of Three-zone to Three-zone_d2.5 AGB models can match majority of the mainstream grains with $\delta^{138}$Ba values > −400 ‰, $\delta^{138}$Ba lower than −400 ‰ can be obtained only with a flatter Zone-II $^{13}$C pocket, whose total mass is between $2.1\times10^{-4}$–$5.3\times10^{-4}$ $M_\odot$ with a $^{13}$C mass fraction of $6.8\times10^{-3}$ (around the ST case). $\delta^{138}$Ba values provide an important constraint for future simulations of the physical mechanism(s) responsible for formation of the $^{13}$C pocket.
3. Zone-II model calculations predict a large enhancement of $^{88}$Sr, in contrast to presolar SiC grain data from Nicolussi et al. (1998). The fact that most grains in that study lie significantly closer to solar values than any of the model predictions indicates that the grains may be highly contaminated. Strontium isotope measurements in acid-cleaned grains can test the degree of strontium contamination in the previous study. Therefore, correlated strontium and barium isotope measurements in acid-cleaned grains will be done in order to better constrain the $^{13}$C profile within the $^{13}$C pocket and the pocket mass. Such measurements will allow us to answer the question: what fraction of the parent AGB stars of mainstream grains has Zone-II-like $^{13}$C pockets?
4. Carbon isotopic ratios in the grains from this study do not require the introduction of the additional extra-mixing processes such as cool bottom processing during the AGB phase. For FRUITY model calculations, RGB extra-mixing needs to be included to better match with the grain data.

5. Based on $\delta^{134}$Ba values of mainstream grains, we derive constraints on the rate of $^{22}Ne(\alpha,n)^{25}Mg$ to be $2.1\times10^{-11}\ cm^3\ mol^{-1}\ s^{-1}$ with an uncertainty of a factor of two, considering the uncertainty of $^{134}$Cs $\beta^-$ decay rate estimated by Goriely (1999). This result is in good agreement with more recently recommended values by Jaeger et al. (2001) and Longland et al. (2012).
6. Negative $\delta^{134}$Ba values cannot be explained by any of the current AGB model calculations. Instead, the negative values may be a signature of *i*-process neutron capture nucleosynthesis during VLTP events in born-again AGB stars. Two mainstream grains with strongly negative $\delta^{134}$Ba values can be explained by VLTP model calculations, indicating nucleosynthesis in parent stars with higher than the *s*-process neutron density. In addition to barium isotopic composition, the carbon isotope ratios of the two grains could be explained by the calculations. In contrast, their silicon isotopic compositions are not consistent with present VLTP model predictions.
7. We reported the first $\delta^{130+132}$Ba values in single mainstream SiC grains and concluded that grains formed before significant $^{135}$Cs decay in the AGB circumstellar envelope, in good agreement with results on grain aggregates. The absence of radiogenic $^{135}$Ba in $\mu m$-size single SiC grains supports condensation of barium atoms into the grains instead of later ion implantation.

Acknowledgements: We thank the anonymous referee for a careful and constructive reading of the manuscript. This work is supported by the NASA Cosmochemistry program, through grants to the University of Chicago and Argonne National Laboratory, by the US Dept. of Energy, BES − Division of Materials Science and Engineering, under contract DEAC02-06CH11357 (CHARISMA facility). NL acknowledges the NASA Earth and Space Sciences Fellowship Program (NNX11AN63H) for support. SC and OS acknowledge support from the Italian Ministry of Education, University and Research under the FIRB2008 program (RBFR08549F-002) and from the PRIN-INAF 2011 project "Multiple populations in Globular Clusters: their role in the Galaxy assembly". SC and OS thank Dr. Luciano Piersanti for continuous scientific valuable discussions on stellar modeling. SB acknowledges financial support from the Joint Institute for Nuclear Astrophysics (JINA, University of Notre Dame, USA) and from Karlsruhe Institute of Technology (KIT, Karlsruhe, Germany). Part of the Torino model numerical calculations has been sustained by B2FH Association (http://www.b2fh.org/). MP & FH acknowledge significant support from NSF grants PHY 02-16783 and PHY 09-22648 (Joint Institute for Nuclear Astrophysics, JINA) and EU MIRG-CT-2006-046520. The continued work on codes and in disseminating data is made possible through funding from NSERC Discovery grant (FH, Canada), and from SNSF an Ambizione grant and the research grant 200020-132816 (MP, Switzerland). MP also thanks the support from EuroGENESIS.

## FIGURE CAPTIONS

Figure 1 The xenon to lanthanum section of the chart of the nuclides. Terrestrial abundances are shown as percentages for stable isotopes (solid squares); laboratory half-lives at room temperature are shown for unstable isotopes (dotted squares). Pure *s*-process nuclides are outlined with thick black squares. The main path of the *s*-process is shown with thick lines; alternative paths (due to branch points) are indicated with thin lines. Neutron-magic nuclides lie on the vertical yellow band at $N = 82$. (A color version of this figure is available in the online journal.)

Figure 2 Three-isotope plots of $\delta(^{134}Ba/^{136}Ba)$, $\delta(^{137}Ba/^{136}Ba)$ and $\delta(^{138}Ba/^{136}Ba)$ versus $\delta(^{135}Ba/^{136}Ba)$ for the 38 grains in this study. Unclassified grains (upside down triangles) are well within the range of mainstream grains and are therefore grouped as mainstream and shown as black dots hereafter. Uncertainties are $\pm 2\sigma$. Dotted lines represent solar barium isotope ratios. (A color version of this figure is available in the online journal.)

Figure 3 Three-isotope plots of $\delta(^{134}Ba/^{136}Ba)$, $\delta(^{137}Ba/^{136}Ba)$and $\delta(^{138}Ba/^{136}Ba)$ versus $\delta(^{135}Ba/^{136}Ba)$ for comparison with previous single mainstream grain data. All grain data are chosen based on the criterion of $2\sigma(\delta^{135}Ba)<160$ ‰. No $\delta(^{134}Ba/^{136}Ba)$ values was reported by Savina et al. (2003a). (A color version of this figure is available in the online journal.)

Figure 4 Three-isotope plots of $\delta(^{137}Ba/^{136}Ba)$ versus $\delta(^{135}Ba/^{136}Ba)$. Linear regression lines (black solid line) of single grain data from this study with 95% confidence (shown as grey region) are compared with that of SiC aggregates (red dashed line) reported in Fig. 1D of Prombo et al. (1993) in (a). Three-zone model calculations in a 2 $M_{\odot}$, 0.5 $Z_{\odot}$ and 3 $M_{\odot}$, 0.5 $Z_{\odot}$ AGB star are shown in (a) & (b) for comparison respectively, with the entire evolution of the AGB envelope composition shown. Symbols plotted only when C>O. (A color version of this figure is available in the online journal.)

Figure 5 Plots of δ($^{134}$Ba/$^{136}$Ba) versus δ($^{135}$Ba/$^{136}$Ba) values. Three-zone AGB model predictions from D3−U2 cases are shown with a range of masses and metallicities constrained by previous studies of mainstream grains. (A color version of this figure is available in the online journal.)

Figure 6 Three-isotope plots of δ($^{134}$Ba/$^{136}$Ba) and δ($^{138}$Ba/$^{136}$Ba) versus δ($^{135}$Ba/$^{136}$Ba). Same grain data as in Fig. 5, are plotted with FRUITY predictions in low mass AGB stars with close-to-solar metallicities. (A color version of this figure is available in the online journal.)

Figure 7 Three-isotope plots of δ($^{138}$Ba/$^{136}$Ba) versus δ($^{135}$Ba/$^{136}$Ba). Same Set of grains as in Figs. 5 & 6 are compared to Three-zone, Three-zone_d2.5, Zone-II & Zone-II_d2.5 model predictions in a 2 $M_{\odot}$, 0.5 $Z_{\odot}$ AGB star. Zone-II model predictions for carbon-rich TPs are shown as filled symbols to distinguish them from Three-zone predictions shown as open symbols. (A color version of this figure is available in the online journal.)

Figure 8 Plots of δ($^{135}$Ba/$^{136}$Ba) versus $^{12}$C/ $^{13}$C. The 24 mainstream SiC grains in Murchison from this study, compared to Zone-II AGB model predictions of a 2 $M_{\odot}$, 0.5 $Z_{\odot}$ AGB star with a range of $^{13}$C efficiency and FRUITY model predictions of 2 $M_{\odot}$ and 3 $M_{\odot}$ AGB stars with close-to-solar metallicity. (A color version of this figure is available in the online journal.)

Figure 9 Plots of δ($^{134}$Ba/$^{136}$Ba) versus δ($^{135}$Ba/$^{136}$Ba). The mainstream SiC grains in Murchison from this study, compared to Zone-II model predictions of a 2 $M_{\odot}$, 0.5 $Z_{\odot}$ AGB star with a range of $^{22}$Ne($\alpha$,$n$)$^{25}$Mg rates. The rate of K94 ($4.14\times10^{-11}$ $cm^3$ $mol^{-1}$ $s^{-1}$ at $T = 3\times10^8$ $K$) refers to the lower limit of recommended $^{22}$Ne($\alpha$,$n$)$^{25}$Mg rate from Käppeler et al. (1994). (A color version of this figure is available in the online journal.)

Figure 10 Three-isotope plots of δ($^{134}$Ba/$^{136}$Ba), δ($^{137}$Ba/$^{136}$Ba) and δ($^{138}$Ba/$^{136}$Ba) versus δ($^{135}$Ba/$^{136}$Ba). Two grains with strongly negative δ$^{134}$Ba are compared with the weighted average barium isotopic composition of a helium-zone ingested with hydrogen after a VLTP during the post-AGB phase (Herwig et al. 2011). The model predictions are shown as lines with different parameter values. The thinner long-dashed lines represent mixing lines of barium isotopic compositions between hydrogen envelope and helium-zone (same respective shade of gray or color) for the two cases with different split times. No *s*-process enhancement is assumed during the AGB phase and the initial inputs are solar, shown as points at zero. (A color version of this figure is available in the online journal.)

Figure 11 Four-isotope plot of δ($^{134}$Ba/$^{136}$Ba) versus $^{12}$C/$^{13}$C. The two grains and model predictions are the same as in Fig. 10. Complete $^{134}$Cs decay is considered for model predictions of δ$^{134}$Ba values. (A color version of this figure is available in the online journal.)

Figure 12 Three-isotope plot of δ($^{130+132}$Ba/$^{136}$Ba) versus δ($^{135}$Ba/$^{136}$Ba) in single mainstream SiC grains and in SiC aggregates. Zone-II model predictions of 2 $M_{\odot}$, 0.5 $Z_{\odot}$ AGB stars are shown as two extreme cases for comparison: no $^{135}$Cs decay and complete $^{135}$Cs decay. For each case, $^{13}$C efficiencies in ST and D3 cases are shown as representative. (A color version of this figure is available in the online journal.)

Table 1. Carbon, Silicon and Barium Grain Data

| Grains | Type | $^{12}C/^{13}C$ | $\delta^{29}Si$ (‰) | $\delta^{30}Si$ (‰) | $\delta^{130+132}Ba$ (‰) | $\delta^{134}Ba$ (‰) | $\delta^{135}Ba$ (‰) | $\delta^{137}Ba$ (‰) | $\delta^{138}Ba$ (‰) |
|---|---|---|---|---|---|---|---|---|---|
| G81 | M | 56±1.0 | 68±11 | 24±13 | | 72±216 | −623±76 | −419±76 | −342±72 |
| G82 | M | 63±1.0 | 26±10 | −3±12 | | 49±244 | −661±82 | −442±86 | −361±80 |
| G94 | M | 58±1.0 | 27±12 | 16±12 | | 24±394 | −610±148 | −363±156 | −345±134 |
| G96 | M | 85±1.0 | −48±12 | −44±12 | | −165±340 | −720±118 | −462±134 | −290±140 |
| G123 | M | 72±1.0 | 68±12 | 57±12 | | 77±216 | −593±104 | −380±82 | −379±68 |
| G125 | M | 71±1.0 | −26±14 | −20±15 | | 144±138 | −575±52 | −345±54 | −308±46 |
| G140 | M | 78±1.0 | 18±13 | 19±14 | | −94±258 | −791±78 | −391±114 | −321±100 |
| G143 | M | 30±0.4 | 97±11 | 85±10 | | 26±376 | −415±186 | −289±210 | −306±164 |
| G146 | M | 75±1.0 | 16±13 | 21±13 | | −68±218 | −714±78 | −457±104 | −447±82 |
| G147 | M | 56±1.0 | 48±16 | 20±18 | −863±54 | 204±104 | −748±20 | −477±30 | −461±24 |
| G160 | M | 95±3.0 | 20±25 | −43±29 | −819±210 | 254±246 | −733±56 | −485±76 | −521±50 |
| G170 | M | 61±1.0 | 44±11 | 39±10 | | −44±320 | −782±80 | −465±124 | −325±108 |
| **G232** | **M** | **94±1.0** | **−32±12** | **−14±12** | | **−606±388** | **−742±158** | **−598±184** | **−477±166** |
| G233 | M | 73±1.0 | 38±26 | 9±29 | | 183±462 | −674±114 | −505±132 | −313±132 |
| G236 | M | 88±1.0 | −4±11 | 29±11 | | 95±292 | −581±144 | −458±100 | −128±124 |
| G243 | M | 82±4.0 | 111±31 | −2±36 | | 102±188 | −617±66 | −433±72 | −345±62 |
| **G244** | **M** | **68±1.0** | **16±9** | **32±11** | | **−433±328** | **−730±138** | **−465±180** | **−330±162** |
| G252 | M | 80±1.0 | 96±10 | 60±13 | | 75±74 | −652±24 | −424±28 | −329±24 |
| G262 | M | 62±1.0 | 81±9 | 50±11 | | 190±204 | −695±52 | −416±72 | −341±58 |
| G265 | M | 97±6.0 | 159±65 | 128±78 | | −289±786 | −813±234 | −523±348 | −639±194 |
| G270 | M | 31±0.4 | 43±8 | 34±10 | | 36±220 | −545±80 | −352±92 | −349±68 |
| G342 | M | 60±1.0 | 98±13 | 56±12 | | −27±252 | −525±88 | −302±106 | −320±82 |
| G372 | M | 66±1.0 | 35±17 | 21±20 | | 200±260 | −651±82 | −421±98 | −440±70 |
| G393 | M | 79±1.0 | 25±8 | 13±11 | | −16±434 | −474±184 | −151±234 | −172±176 |
| G77 | U[a] | | | | | 51±282 | −699±88 | −451±98 | −256±106 |
| G86 | U | | | | −881±84 | 19±108 | −787±26 | −494±34 | −366±36 |
| G87 | U | | | | | 324±274 | −690±64 | −407±86 | −338±78 |
| G89 | U | | | | | −101±244 | −168±166 | −90±138 | −190±108 |
| G90 | U | | | | | −99±570 | −863±126 | −552±190 | −387±198 |
| G130 | U | | | | | 342±484 | −526±188 | −409±162 | −197±170 |
| G132 | U | | | | −913±72 | 176±114 | −751±28 | −486±36 | −483±28 |
| G148 | U | | | | −714±182 | 192±172 | −685±44 | −396±62 | −481±40 |
| G175 | U | | | | | −205±604 | −728±192 | −598±214 | −360±218 |
| G224 | U | | | | | 152±158 | −518±48 | −351±56 | −341±44 |
| G245 | U | | | | −780±90 | 106±68 | −741±18 | −462±24 | −396±20 |
| G260 | U | | | | −774±96 | 180±120 | −664±32 | −403±42 | −557±24 |
| G379 | U | | | | | 128±68 | −47±42 | −56±38 | −97±30 |
| G339 | U | | | | | 32±302 | −527±102 | −306±122 | −352±90 |

**Notes**: Uncertainties are given as 2σ. Two grains with negative $\delta^{134}Ba$ values within 2σ errors are highlighted.

[a]: U stands for unclassified.

Table 2. Three-zone $^{13}$C Pocket in the ST Case of Torino Postprocessing AGB Models

| ST $^{13}$C pocket | Zone-I (Innermost) | Zone-II (Middle) | Zone-III (Outermost) |
|---|---|---|---|
| Total M ($M_\odot$) | $4.00\times10^{-4}$ | $5.30\times10^{-4}$ | $7.50\times10^{-6}$ |
| $X$ ($^{13}$C) | $3.20\times10^{-3}$ | $6.80\times10^{-3}$ | $1.60\times10^{-2}$ |
| $X$ ($^{14}$N) | $1.07\times10^{-4}$ | $2.08\times10^{-4}$ | $2.08\times10^{-3}$ |

Table 3. The Dependence of $\delta^{138}$Ba Values on $^{13}$C Pocket Internal Structures in ST Case

| Models | Total mass ($\times 10^{-4} M_{\odot}$) | $\delta^{138}$Ba (‰) (C = O) | $\delta^{138}$Ba (‰) (Minimum) | $\delta^{138}$Ba (‰) (Last TP) |
|---|---|---|---|---|
| Zone-II _u2 | 10.6 | −310 | −358 | −150 |
| Zone-II | 5.30 | −426 | −426 | −376 |
| Zone-II_d1.5 | 3.53 | −447 | −408 | −456 |
| Zone-II_d2 | 2.65 | −444 | −502 | −491 |
| Zone-II_d2.5 | 2.12 | −434 | −514 | −509 |
| Three-zone_u2 | 18.76 | −102 | −253 | 1 |
| Three-zone | 9.38 | −274 | −274 | −185 |
| Three-zone_d2 | 4.69 | −326 | −336 | −308 |
| Three-zone_d2.5 | 3.75 | −322 | −352 | −345 |
| Three-zone_d4 | 2.34 | −322 | −352 | −345 |

Table 4. Index Values from Torino Model Calculations with a Zone-II $^{13}C$ Pocket

| Cases | $\tau_0$ (mbarn$^{-1}$) | [hs/ls] | [ls/Fe] | [hs/Fe] | [Pb/Fe] |
|---|---|---|---|---|---|
| U2 | 0.51 | 0.62 | 1.39 | 2.01 | 1.86 |
| U1.3 | 0.39 | −0.12 | 1.61 | 1.49 | 1.54 |
| ST | 0.32 | −0.40 | 1.61 | 1.21 | 1.22 |
| D1.5 | 0.24 | −0.46 | 1.32 | 0.86 | 0.49 |
| D2 | 0.19 | −0.50 | 1.04 | 0.54 | 0.19 |
| D3 | 0.15 | −0.50 | 0.69 | 0.19 | 0.03 |

**Note**: Calculations are for a 2 $M_{\odot}$, 0.5 $Z_{\odot}$ AGB star.

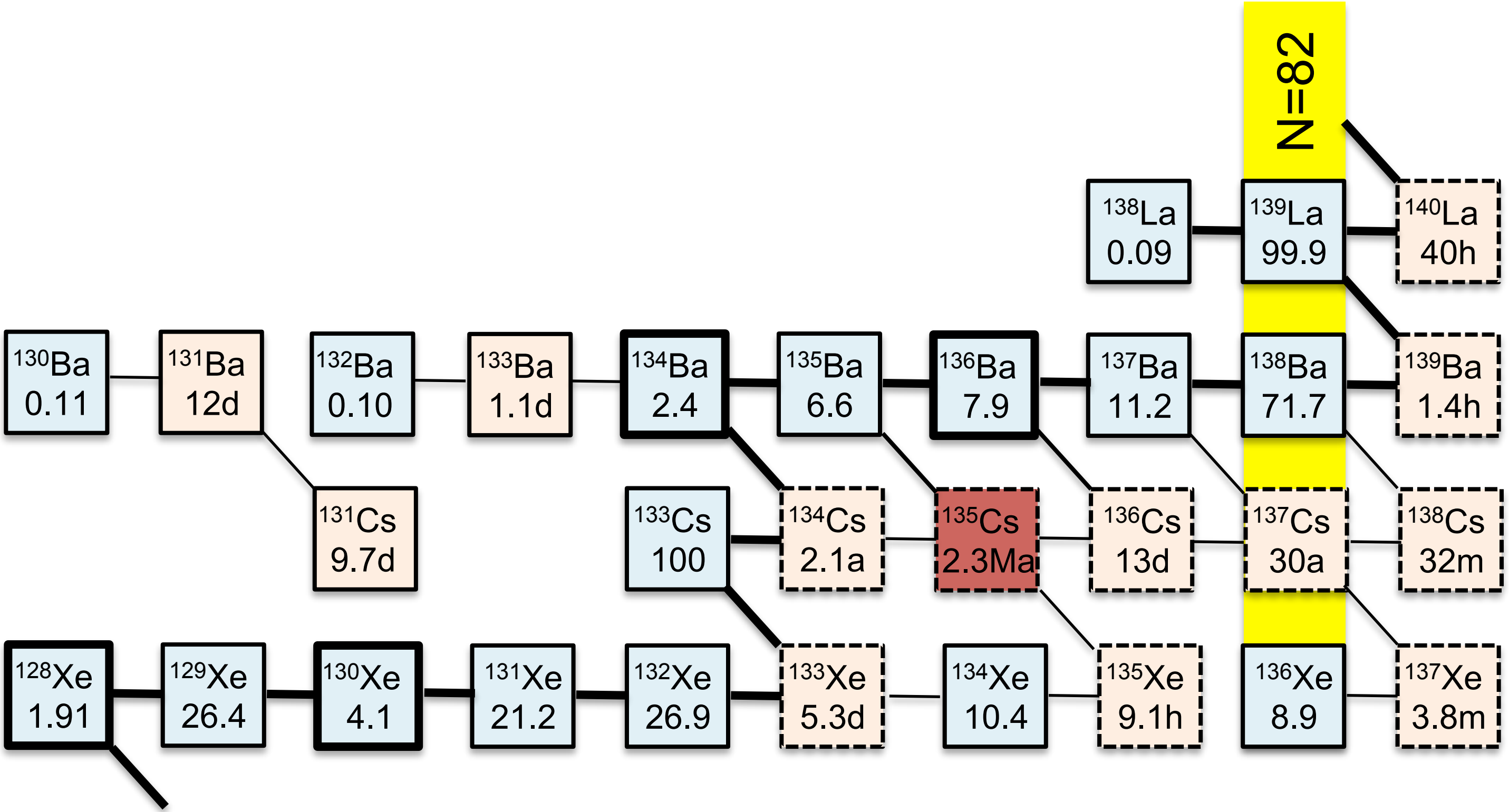

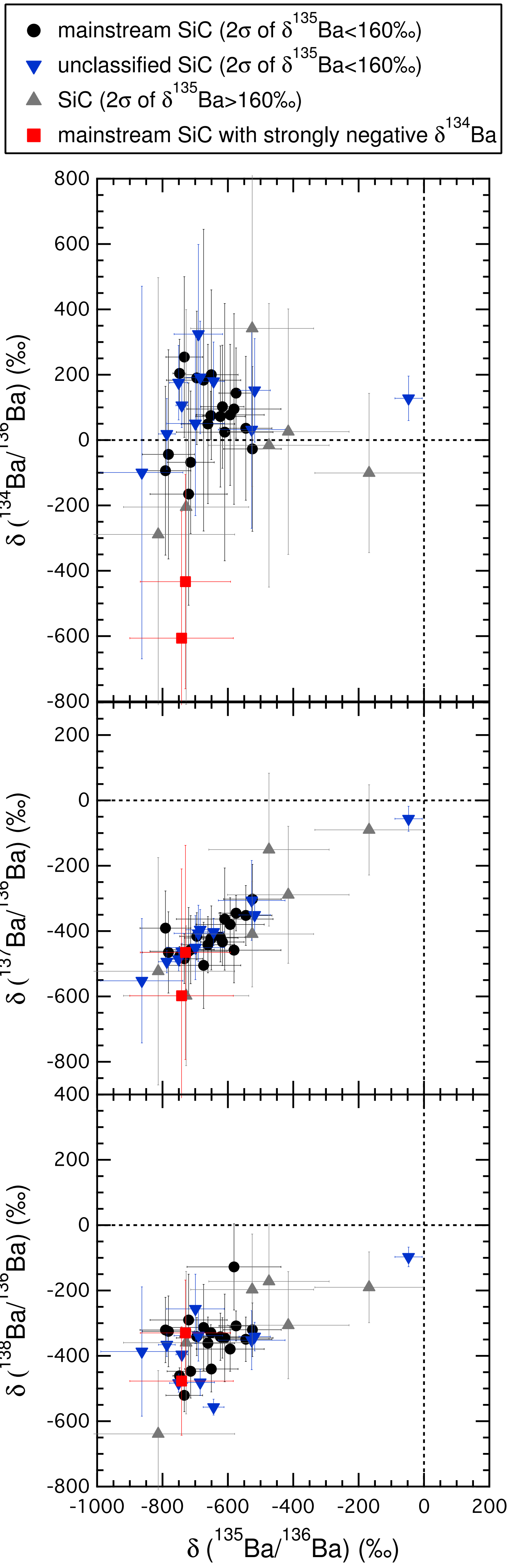

mainstream SiC (2σ of δ135Ba<160‰)
unclassified SiC (2σ of δ135Ba<160‰)
SiC (2σ of δ135Ba>160‰)
mainstream SiC with strongly negative δ134Ba
δ (134Ba/136Ba) (‰)
δ (137Ba/136Ba) (‰)
δ (138Ba/136Ba) (‰)
δ (135Ba/136Ba) (‰)
800
600
400
200
0
-200
-400
-600
-800
-1000
200

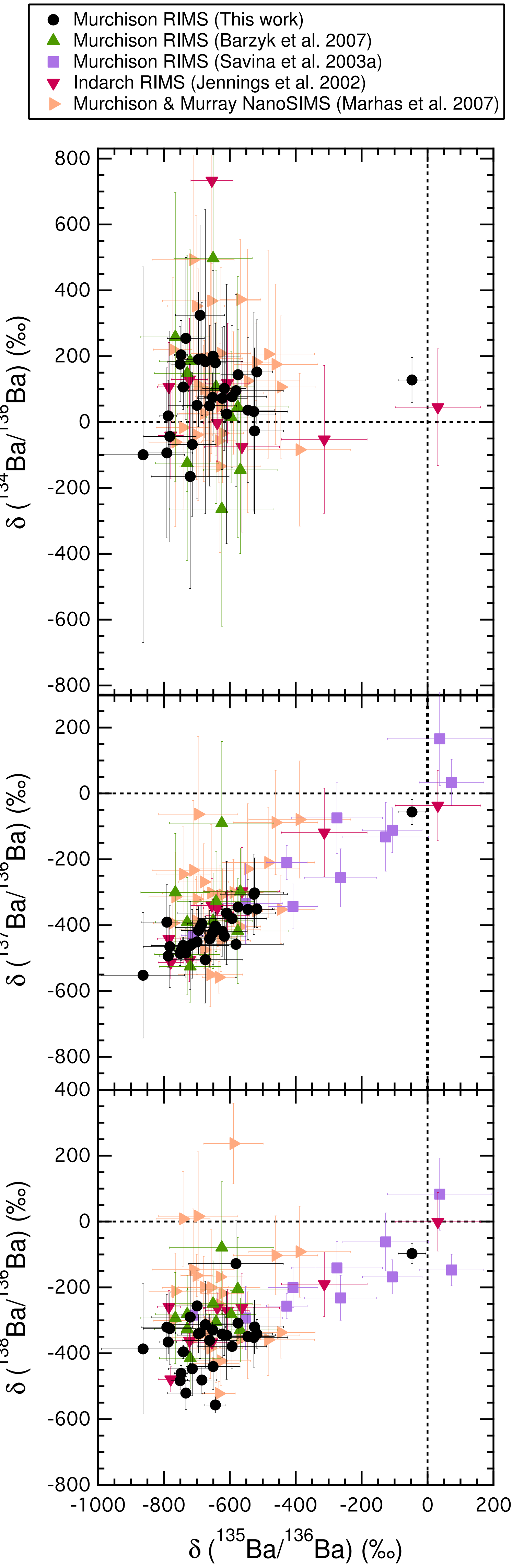

Murchison RIMS (This work)
Murchison RIMS (Barzyk et al. 2007)
Murchison RIMS (Savina et al. 2003a)
Indarch RIMS (Jennings et al. 2002)
Murchison & Murray NanoSIMS (Marhas et al. 2007)
δ (134Ba/136Ba) (‰)
δ (137Ba/136Ba) (‰)
δ (138Ba/136Ba) (‰)
δ (135Ba/136Ba) (‰)

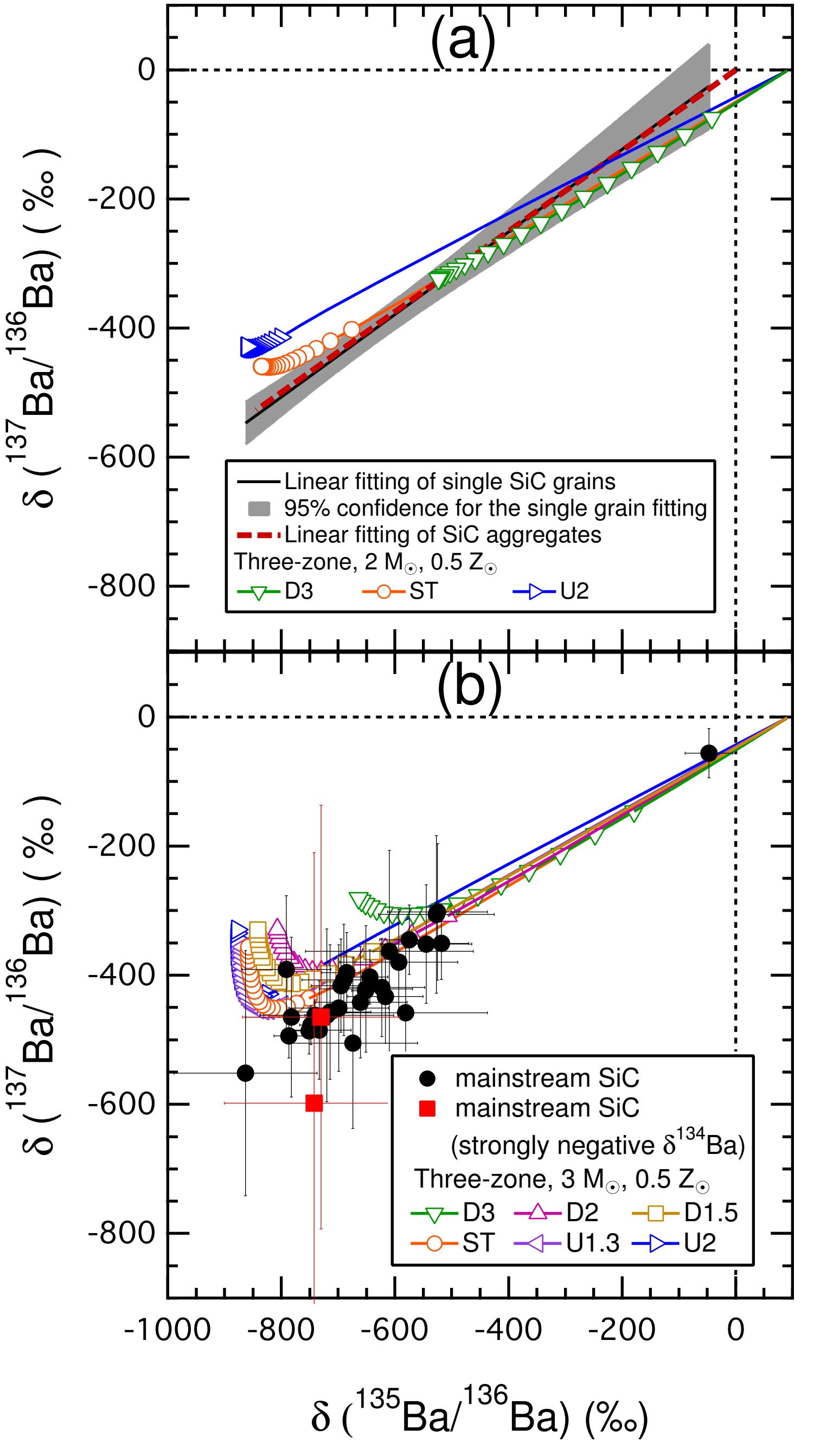

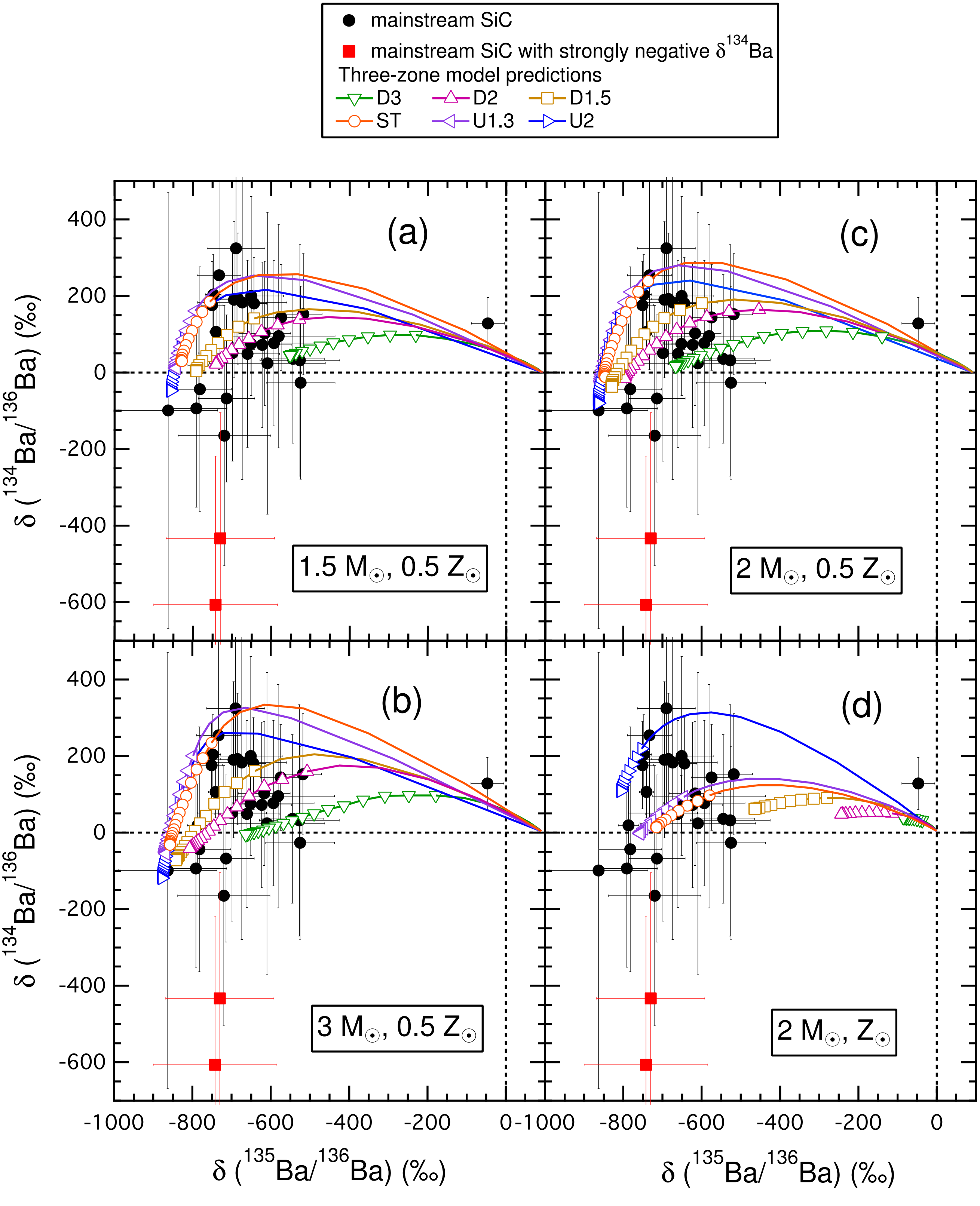

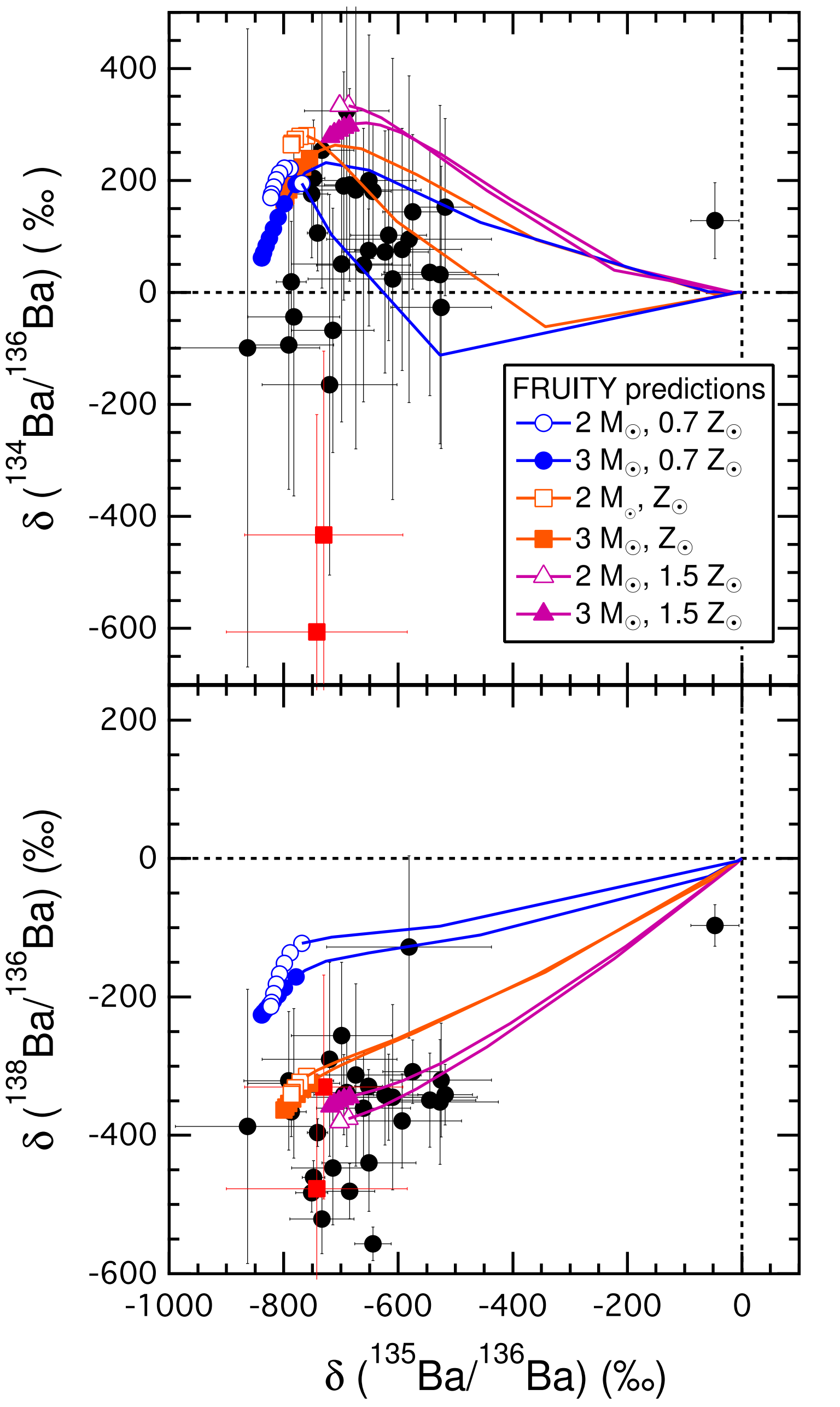

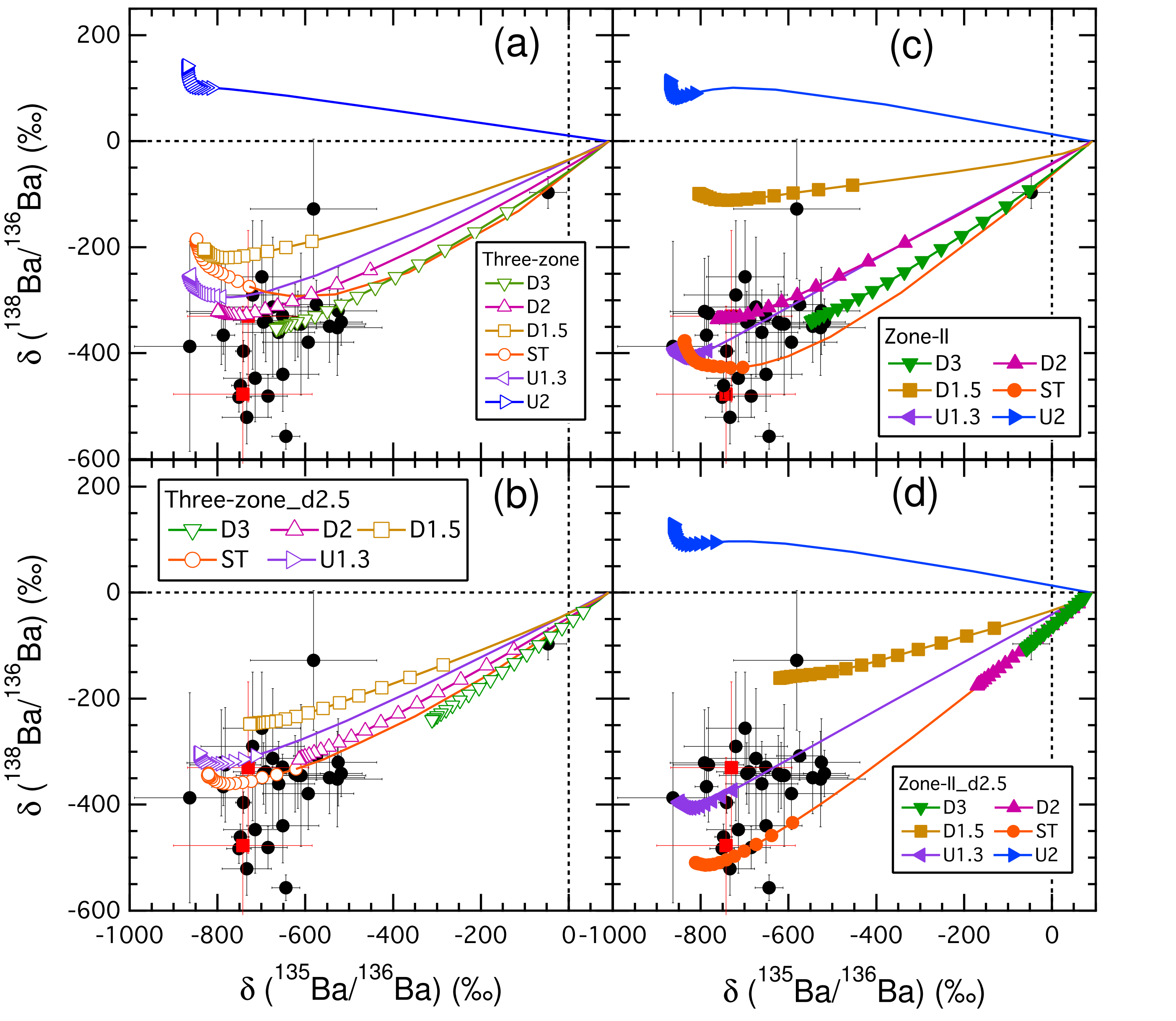

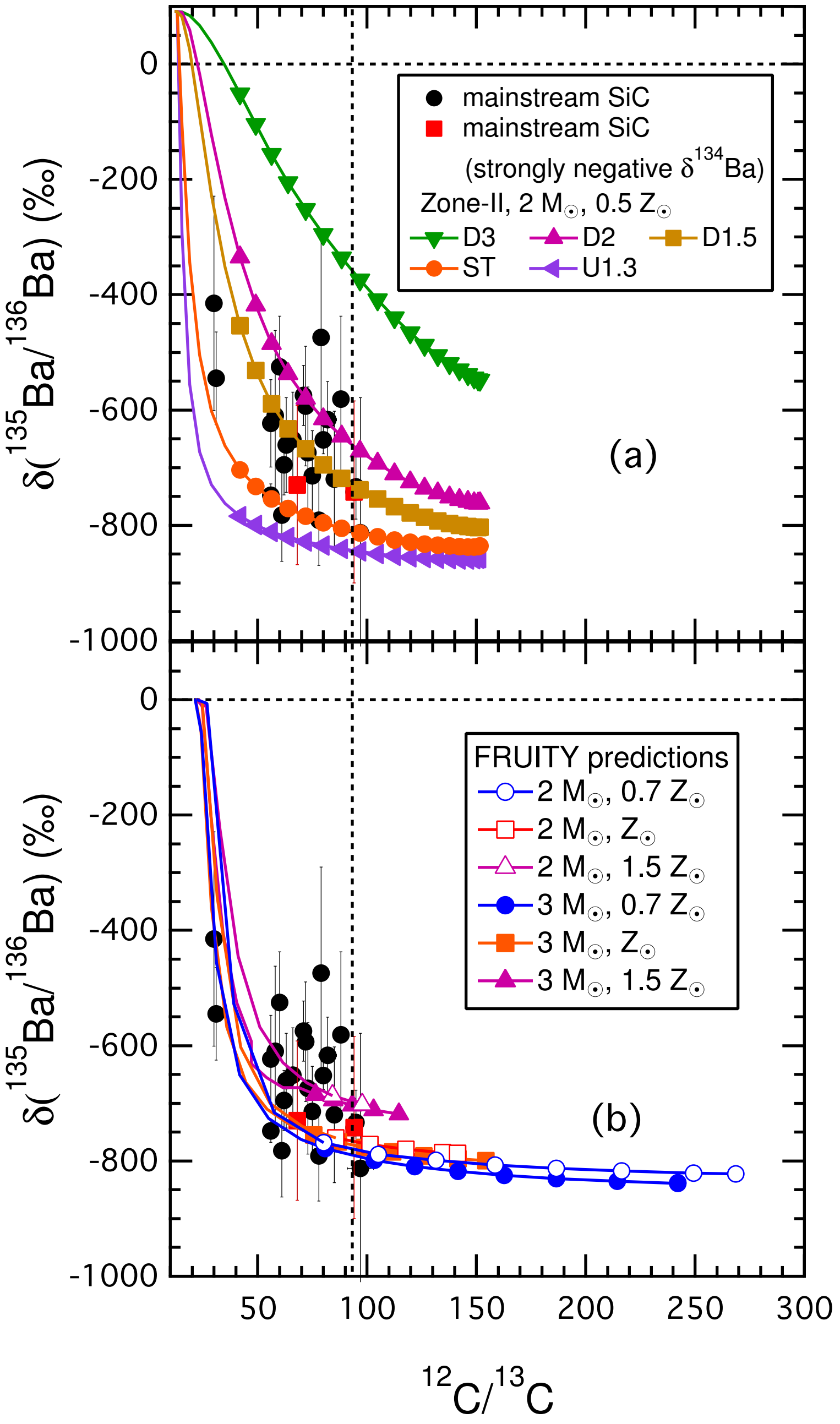

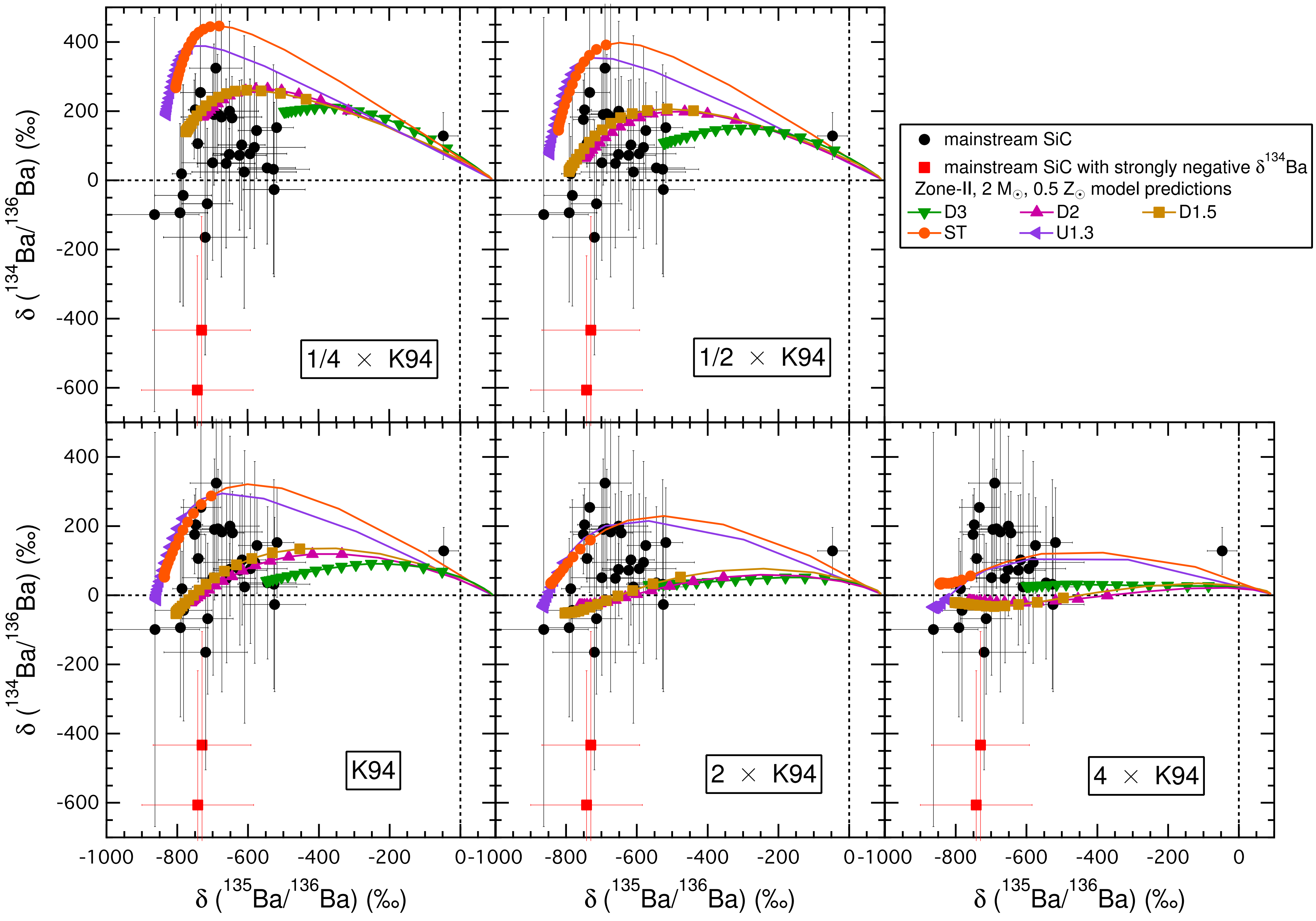

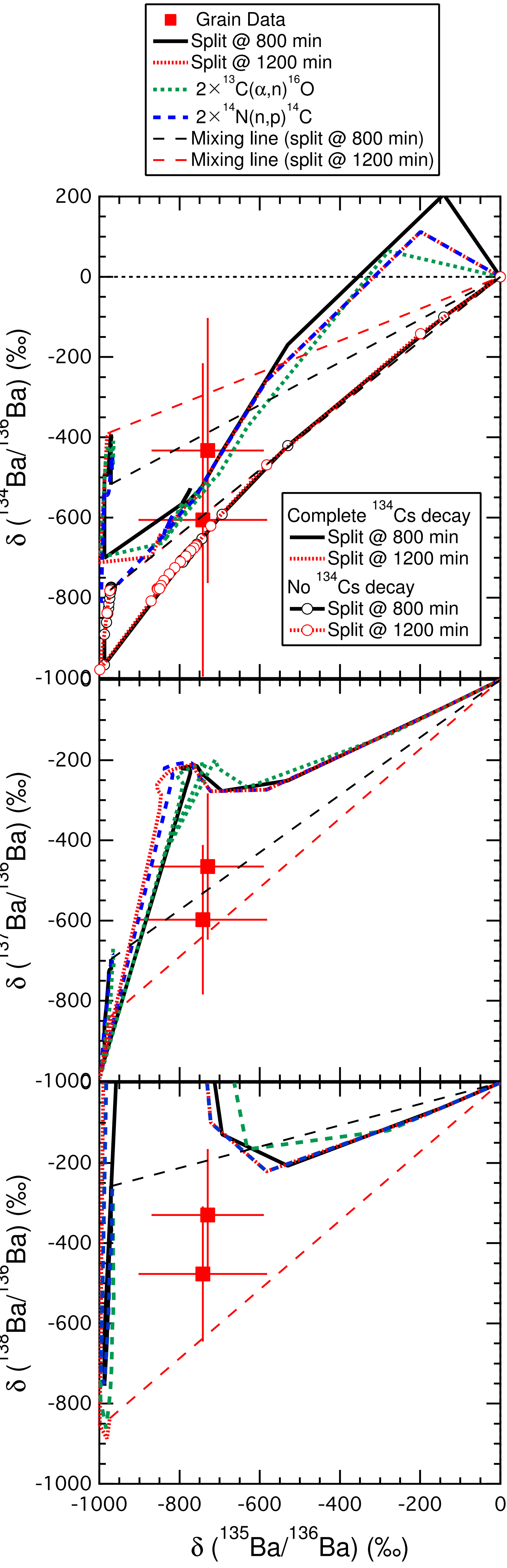

Grain Data
Split @ 800 min
Split @ 1200 min
2× $^{13}$C(α,n)$^{16}$O
2× $^{14}$N(n,p)$^{14}$C
Mixing line (split @ 800 min)
Mixing line (split @ 1200 min)
Complete $^{134}$Cs decay
Split @ 800 min
Split @ 1200 min
No $^{134}$Cs decay
Split @ 800 min
Split @ 1200 min
δ ($^{134}$Ba/$^{136}$Ba) (‰)
δ ($^{137}$Ba/$^{136}$Ba) (‰)
δ ($^{138}$Ba/$^{136}$Ba) (‰)
δ ($^{135}$Ba/$^{136}$Ba) (‰)
200
0
-200
-400
-600
-800
-1000
-1000
-800
-600
-400
-200
0

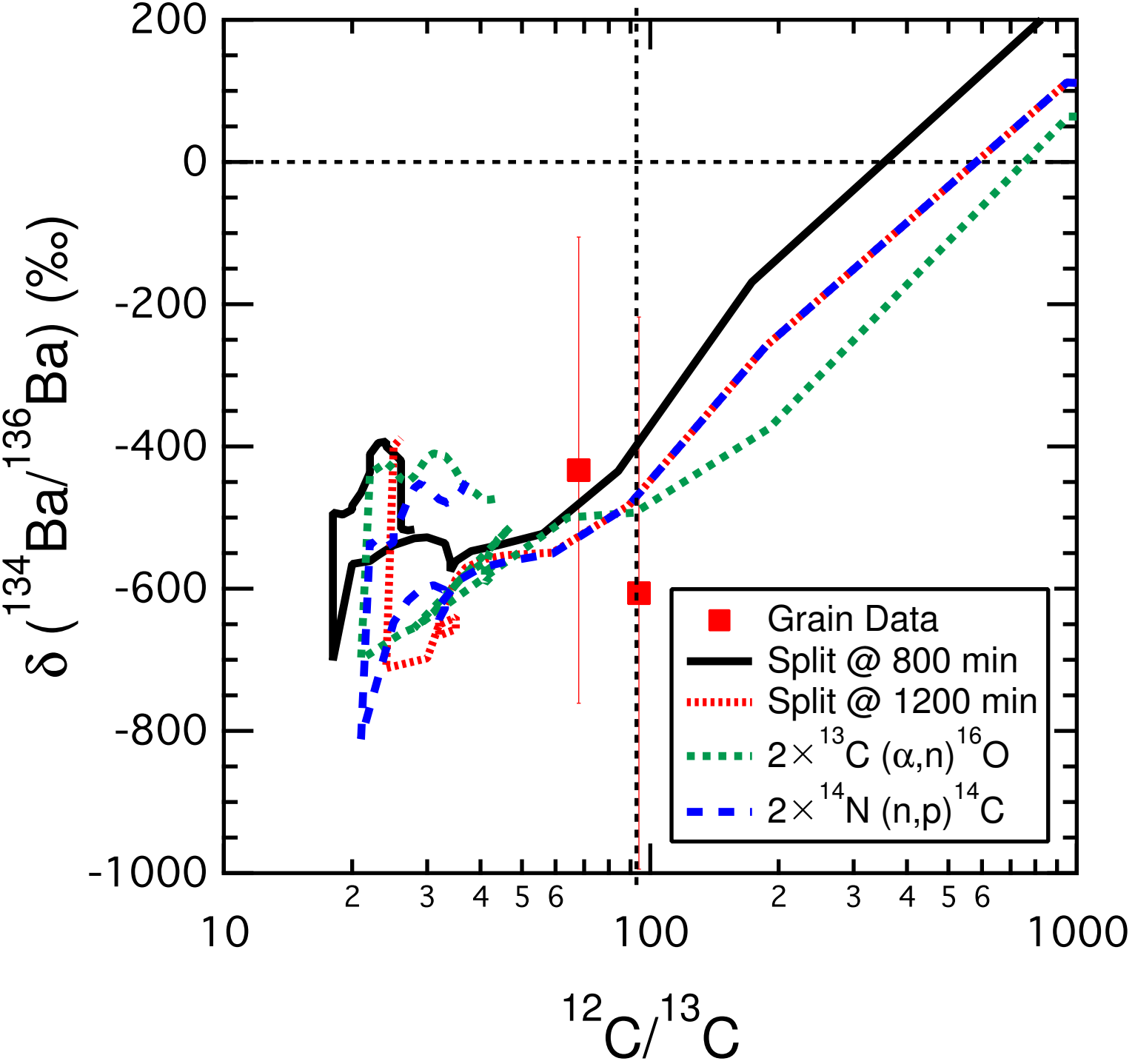

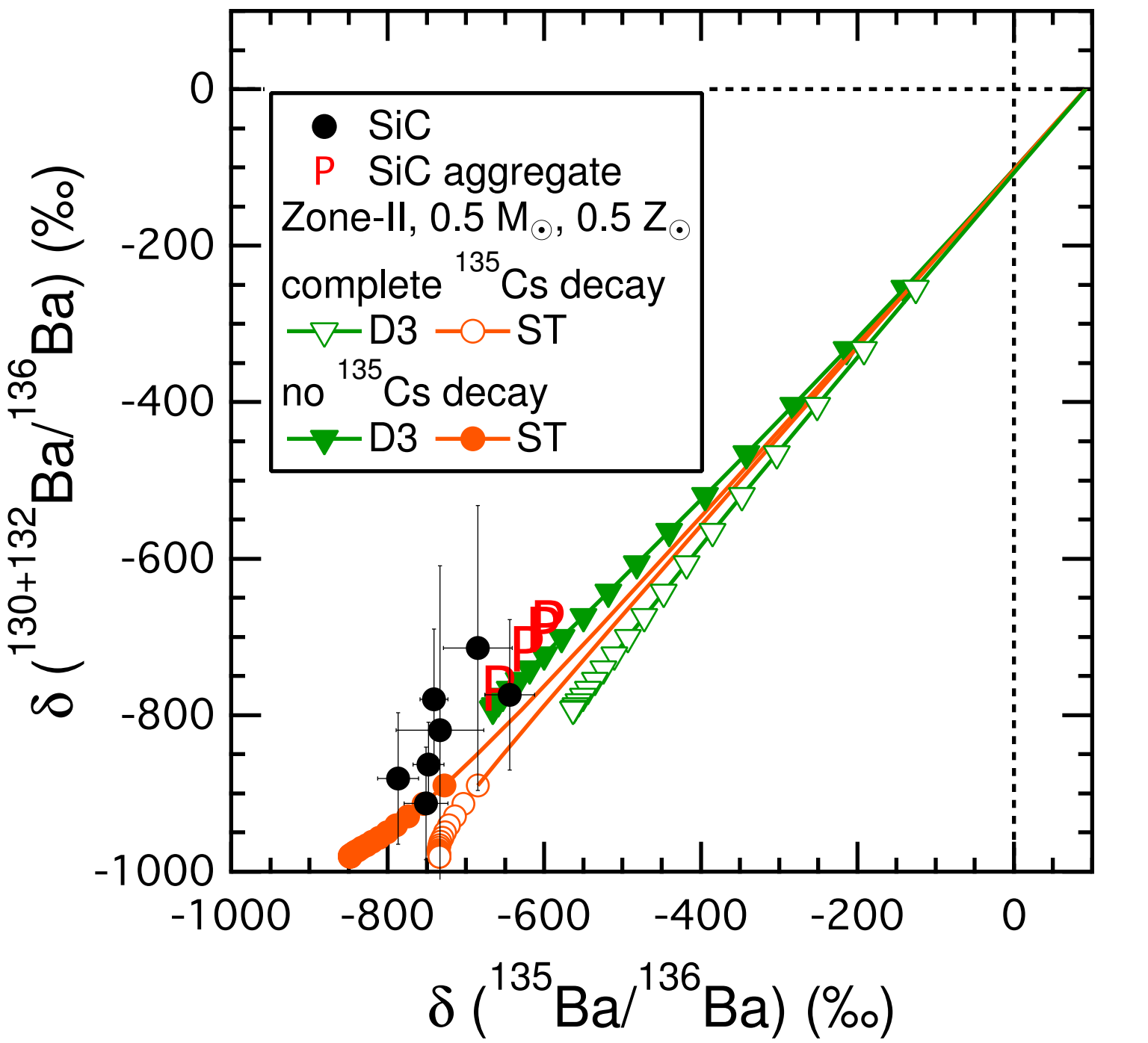